\documentclass[graybox,footinfo]{svmult}

\smartqed
\usepackage{mathptmx}       
\usepackage{helvet}         
\usepackage{courier}        
\usepackage{type1cm}        
\usepackage{graphicx}       

\usepackage{array,colortbl}
\usepackage{amsmath,amsfonts,amssymb,bm} 
\DeclareFontFamily{U}{mathx}{\hyphenchar\font45}
\DeclareFontShape{U}{mathx}{m}{n}{
      <5> <6> <7> <8> <9> <10>
      <10.95> <12> <14.4> <17.28> <20.74> <24.88>
      mathx10
      }{}
\DeclareSymbolFont{mathx}{U}{mathx}{m}{n}
\DeclareFontSubstitution{U}{mathx}{m}{n}
\DeclareMathAccent{\widecheck}      {0}{mathx}{"71}

\renewcommand{\email}[1]{\emailname: #1} 

\usepackage{mathptmx}       
\usepackage{helvet}         
\usepackage{courier}        

\usepackage{makeidx}         
\usepackage{graphicx}        
\usepackage[bottom]{footmisc}

\usepackage{latexsym}
\usepackage{amsmath}
\usepackage{amsfonts}
\usepackage{amssymb}
\usepackage{bm}

\usepackage{url}
\usepackage{algorithm}
\usepackage{algorithmic}
\usepackage[misc,geometry]{ifsym}

\spdefaulttheorem{assumption}{Assumption}{\upshape \bfseries}{\itshape}
\spdefaulttheorem{algo}{Algorithm}{\upshape \bfseries}{\itshape}

\newcommand{\Q}{{\mathbb{Q}}} 
\newcommand{\R}{{\mathbb{R}}} 

\DeclareSymbolFont{bbold}{U}{bbold}{m}{n}
\DeclareSymbolFontAlphabet{\mathbbold}{bbold}
\newcommand{\ind}{{\mathbbold{1}}}

\newcommand{\setX}{{\mathfrak{X}}}

\DeclareSymbolFont{bbold}{U}{bbold}{m}{n}
\DeclareSymbolFontAlphabet{\mathbbold}{bbold}

\usepackage{microtype} 

\usepackage[colorlinks=true,linkcolor=black,citecolor=black,urlcolor=black]{hyperref}
\usepackage{bookmark}
\makeatletter 
  \providecommand*{\toclevel@author}{999}
  \providecommand*{\toclevel@title}{0}
\makeatother

\usepackage[english]{babel}

\begin{document}

\newcommand{\Rp}{\mathbb{R}^+}
\newcommand{\dd}{\mathrm{d}}
\newcommand{\dx}{\dd x}
\newcommand{\bigO}{\mathcal{O}}
\newcommand{\Pm}{\mathbb{P}} 
\newcommand{\M}{\mathbb{M}}
\newcommand{\Qh}{\widehat{\mathbb{Q}}} 
\newcommand{\Qb}{\overline{\mathbb{Q}}} 
\newcommand{\Unif}{\mathcal{U}} 

\title*{Sequential quasi-Monte Carlo: Introduction for Non-Experts, Dimension
Reduction, Application to Partly Observed Diffusion Processes}
\titlerunning{SQMC: introduction, dimension
reduction and application to diffusion processes} 
\author{Nicolas Chopin \and Mathieu Gerber}
\institute{
Nicolas Chopin 
\at CREST-ENSAE, 3 Av. Pierre Larousse, 92245 Malakoff, France.
\email{nicolas.chopin@ensae.fr}
\and 
Mathieu Gerber 
\at School of Mathematics, University of Bristol, University Walk, Clifton, Bristol BS8 1TW, UK.
\email{mathieu.gerber@bristol.ac.uk}
}
\maketitle

\abstract{
SMC (Sequential Monte Carlo) is a class of Monte Carlo algorithms for
filtering and related sequential problems. \cite{MR3351446} introduced
SQMC (Sequential quasi-Monte Carlo), a QMC version of SMC. This paper
has two objectives: (a) to introduce Sequential Monte Carlo to the
QMC community, whose members are usually less familiar with state-space
models and particle filtering; (b) to extend SQMC to the filtering
of continuous-time state-space models, where the latent process is
a diffusion. A recurring point in the paper will be the notion of
dimension reduction, that is how to implement SQMC in such a way that
it provides good performance despite the high dimension of the problem. 
}

\section{Introduction}

SMC (Sequential Monte Carlo) is a class of algorithms that provide
Monte Carlo approximations of a sequence of distributions. The main
application of SMC is the filtering problem: a phenomenon of interest
is modelled as a Markov chain $\left\{ X_{t}\right\}$, which is
not observed directly; instead one collects sequentially data such
as e.g. $Y_{t}=f(X_{t})+V_{t}$, where $V_{t}$ is a noise term.
Filtering amounts to computing the distribution of $X_{t}$ given
$Y_{0:t}=(Y_{0},\ldots,Y_{t})$, the data collected up to time $t$.
Filtering and related problems play an important role in target tracking
(where $X_{t}$ is the position of the target, say a ship), robotic
mapping (where $X_{t}$ is the position of the robot), Epidemiology
(where $X_{t}$ is e.g. the number of infected cases), Finance ($X_{t}$
is the volatility of a given asset) and many other fields. See e.g.
the book of \cite{DouFreiGor}. 

In \cite{MR3351446}, we introduced SQMC (Sequential quasi-Monte
Carlo), a QMC version of Sequential Monte Carlo. As other types of
QMC algorithms, the main advantage of SQMC is the better rate of convergence
one may expect, relative to SMC methods. 

It is difficult to write a paper that bridges the gap between two
scientific communities; in this case, QMC experts on one side, and
Statisticians working on Monte Carlo methods (MCMC and SMC) on the
other side. We realise now that \cite{MR3351446} may be more approachable
by the latter than by the former. In particular, that paper spends
time explaining basic QMC notions to non-experts, but it does not
do the same for SMC. 

To address this short-coming, and hopefully generate some interest
about SQMC in the QMC community, we decided to devote the first part
of this paper to  introducing the motivation and basic
principles of SMC. We do so using the so-called Feynman-Kac formalism,
which is deemed to be abstract, but may be actually more approachable
to non-Statisticians. 

The second part of this paper discusses how to extend SQMC to the
filtering of continuous-time state-space models; i.e. models where
the underlying signal is e.g. a diffusion process. These models are popular in
Finance and in Biology. What makes this extension interesting is that the
inherent dimension of such  models is infinity, whereas the performance of SQMC
seems to deteriorate with the dimension (according to the  numerical studies in
\cite{MR3351446}). However, by using the Markov property of the latent process,
we are able to make some parts of SQMC operate in a low dimension, and, as
result, to make it perform well (and significantly better than SMC) despite 
the infinite dimension of the problem. 

\section{SMC}

\subsection{Basic Notions and Definitions}

The state space $\setX$ of interest in the paper is always an open
subset of $\R^{d}$, which we equip with the Lebesgue measure.

We use the standard colon short-hand for collections of random variables
and related quantities: e.g. $Y_{0:t}$ denote $Y_{0},\ldots,Y_{t}$,
$X_{t}^{1:N}$ denote $X_{t}^{1},\ldots,X_{t}^{N}$, and so on. When
such variables are vectors, we denote by $X_{t}(k)$ their $k-$th
component. 

\subsection{Feynman-Kac Formalism}

The phrase `Feynman-Kac model' comes from Probability theory, where
`model' means distributions for variables of interest, and not specifically
observed variables (i.e. data, as in Statistics). A Feynman-Kac model
consists of: 
\begin{enumerate}
\item The law of a (discrete-time) Markov process $\left\{ X_{t}\right\} $,
specified through an initial distribution $\M_{0}(\dx_{0})$, and
a sequence of Markov kernels $M_{t}(x_{t-1},\dx_{t})$; i.e. $M_{t}(x_{t-1},\dx_{t})$
is the distribution of $X_{t}$, conditional on $X_{t-1}=x_{t-1}$; 
\item A sequence of so-called potential (measurable) functions, $G_{0}:\setX\rightarrow\Rp$,
	$G_{t}:\setX\times\setX\rightarrow\Rp$. ($\Rp=[0,+\infty)$.)  
\end{enumerate}
From these objects, one defines the following sequence of probability
distributions:

\[
\Q_{t}(\dx_{0:t})=\frac{1}{L_{t}}\left\{ G_{0}(x_{0})\prod_{s=1}^{t}G_{s}(x_{s-1},x_{s})\right\} \M_{0}(\dx_{0})\prod_{s=1}^{t}M_{s}(x_{s-1},\dx_{s})
\]
where $L_{t}$ is simply the normalising constant: 
\[
L_{t}=\int_{\setX^{T+1}}\left\{ G_{0}(x_{0})\prod_{s=1}^{t}G_{s}(x_{s-1},x_{s})\right\} \M_{0}(\dx_{0})\prod_{s=1}^{t}M_{s}(x_{s-1},\dx_{s}).
\]
(We assume that $0<L_{t}<+\infty.$) A good way to think of Feynman-Kac
models is that of a sequential change of measure, from the law of
the Markov process $\left\{ X_{t}\right\} $, to some modified law
$\Q_{t}$, where the modification applied at time $t$ is given by
function $G_{t}$. In computational terms, one can also think of (sequential)
importance sampling: we would like to approximate $\Q_{t}$ by simulating
process $\left\{ X_{t}\right\} $, and re-weight realisations at time
$t$ by function $G_{t}$. Unfortunately the performance of this basic
approach would quickly deteriorate with time. 
\begin{example}
\label{exa:rare-event}Consider a Gaussian auto-regressive process,
$X_{0}\sim N(0,1),$ $X_{t}=\phi X_{t-1}+V_{t}$, $V_{t}\sim N(0,1)$,
for $t\geq1$, and take $G_{t}(x_{t-1},x_{t})=\ind_{\Rp}(x_{t})$.
Then, if we use sequential importance sampling, the number of simulated
trajectories that would get a non-zero weight would decrease quickly
with time. In particular, the probability of `survival' at time $t$
would be $2^{-(t+1)}$ for $\phi=0$. 
\end{example}
The successive distributions $\Q_{t}$ are related as follows: 

\begin{equation}
\Q_{t}(\dx_{0:t})=\frac{1}{\ell_{t}}\Q_{t-1}(\dx_{0:t-1})M_{t}(x_{t-1},\dx_{t})G_{t}(x_{t-1},x_{t})\label{eq:recursion_joint}
\end{equation}
where $\ell_{t}=L_{t}/L_{t-1}$. There are many practical settings
(as discussed in the next section) where one is interested only in
approximating the \emph{marginal} distribution $\Q_{t}(\dx_{t})$,
i.e. the marginal distribution of variable $X_{t}$ relative to the
joint distribution $\Q_{t}(\dx_{0:t})$. One can deduce from (\ref{eq:recursion_joint})
the following recursion for these marginals: $\Q_{t}(\dx_{t})$ is
the marginal distribution of variable $X_{t}$ with respect to the bi-variate
distribution
\begin{equation}
\Q_{t}(\dx_{t-1:t})=\frac{1}{\ell_{t}}\Q_{t-1}(\dx_{t-1})M_{t}(x_{t-1},\dx_{t})G_{t}(x_{t-1},x_{t}).\label{eq:forward_recursion}
\end{equation}
Note the dramatic dimension reduction: the initial definition of $\Q_{t}$
involved integrals with respect to $\setX^{t+1}$, but with the above
recursion one may obtain expectations with respect to $\Q_{t}(\dx_{t})$
by computing $t+1$ integrals with respect to $\setX^{2}$. 

\subsection{Feynman-Kac in Practice\label{subsec:Feynman-Kac-in-practice}}

The main application of the Feynman-Kac formalism is the filtering
of a state-space model (also known as a hidden Markov model). This
time, `model' has its standard (statistical) meaning, i.e. a probability
distribution for observed data. 

A state-space model involves two discrete-time processes $\left\{ X_{t}\right\} $
and $\left\{ Y_{t}\right\} $; $\{X_{t}\}$ is Markov, and unobserved,
$\left\{ Y_{t}\right\} $ is observed, and is such that variable $Y_{t}$
conditional on $X_{t}$ and all $(X_{s},Y_{s})$, $s\neq t$ depends
only on $X_{t}$. The standard way to specify this model is through: 
\begin{enumerate}
\item The initial distribution $\Pm_{0}(\dx_{0})$ and the Markov kernels
$P_{t}(x_{t-1},\dx_{t})$ that define the law of the process $\left\{ X_{t}\right\} $; 
\item The probability density $f_{t}(y_{t}|x_{t})$ of $Y_{t}|X_{t}=x_{t}$. 
\end{enumerate}
\begin{example}
\label{exa:StochVol}The stochastic volatility model is a state-space
model popular in Finance (e.g., \cite{kim1998}). One observes the
log-return $Y_{t}$ of a given asset, which is distributed according
to $Y_{t}|X_{t}=x_{t}\sim N(0,e^{x_{t}})$. The quantity $X_{t}$
represents the (unobserved) market volatility, and evolves according
to an auto-regressive process: 
\[
X_{t}-\mu=\phi(X_{t-1}-\mu)+\sigma V_{t},\qquad V_{t}\sim N(0,1).
\]
For $X_{0}$, one may take $X_{0}\sim N\left(\mu,\sigma^{2}/(1-\phi^{2})\right)$
to make the process $\left\{ X_{t}\right\} $ stationary. 
\end{example}
 
\begin{example}
The bearings-only model is a basic model in target tracking, where
$X_{t}$ represents the current position (in $\R^{2}$) of a target,
and $Y_{t}$ is a noisy angular measurement obtained by some device
(such as a radar): 
\[
Y_{t}=\arctan\left(\frac{X_{t}(2)}{X_{t}(1)}\right)+V_{t},\quad V_{t}\sim N(0,\sigma^{2}), 
\]
 where $X_{t}(1)$, $X_{t}(2)$ denote the two components of vector
$X_{t}$. There are several standard ways to model the motion of the
target; the most basic one is that of a random walk. See e.g. \cite{arulampalam2002tutorial}
for more background on target tracking.
\end{example}
Filtering is the task of computing the distribution of variable $X_{t}$,
conditional on the data acquired until time $t$, $Y_{0:t}$. It is
easy to check that, by taking a Feynman-Kac model such that 
\begin{itemize}
\item the process $\{X_{t}\}$ has the same distribution as in the considered
model; i.e. $\M_{0}(\dx_{0})=\Pm_{0}(\dx_{0}),$ $M_{t}(x_{t-1},\dx_{t})=P_{t}(x_{t-1},\dx_{t})$
for any $x_{t-1}\in\setX$;
\item the potential functions are set to $G_{t}(x_{t-1},x_{t})=f_{t}(y_{t}|x_{t})$;
\end{itemize}
then one recovers as $\Q_{t}(\dx_{0:t})$ the distribution of variables
$X_{0:t}$, conditional on $Y_{0:t}=y_{0:t}$; in particular $\Q_{t}(\dx_{t})$
is the filtering distribution of the model. 

We call this particular Feynman-Kac representation of the filtering problem
the bootstrap model. Consider now a Feynman-Kac model with an\emph{
arbitrary }distribution for the Markov process $\{X_{t}\}$, and with
potential 
\[
G_{t}(x_{t-1},x_{t})=\frac{P_{t}(x_{t-1},\dx_{t})f_{t}(y_{t}|x_{t})}{M_{t}(x_{t-1},\dx_{t})}, 
\]
the Radon-Nikodym derivative of $P_{t}(x_{t-1},\dx_{t})f_{t}(y_{t}|x_{t})$
with respect to $M_{t}(x_{t-1},\dx_{t})$ (assuming the latter dominates
the former). Whenever kernels $P_{t}$ and $M_{t}$ admit conditional
probability densities (with respect to a common dominating measure),
this expression simplifies to:
\begin{equation}
G_{t}(x_{t-1},x_{t})=\frac{p_{t}(x_{t}|x_{t-1})f_{t}(y_{t}|x_{t})}{m_{t}(x_{t}|x_{t-1})}.\label{eq:guided_Gt}
\end{equation}

Then again it is a simple exercise to check that one recovers as $\Q_{t}(\dx_{t})$
the filtering distribution of the considered model. We call any Feynman-Kac
model of this form a \emph{guided} model. The bootstrap model corresponds
to the special case where $P_{t}=M_{t}$. 

We shall see in the following section that each Feynman-Kac model
generates a different SMC algorithm. Thus, for a given state-space
model, we have potentially an infinite number of SMC algorithms that
may be used to approximate its sequence of filtering distributions.
Which one to choose? We return to this point in Section \ref{subsec:Back-to-SSMs}. 

\subsection{Sequential Monte Carlo\label{subsec:SMC}}

Consider a given Feynman-Kac model. Sequential Monte Carlo amounts
to compute recursive Monte Carlo approximations to the marginal distributions
$\Q_{t}(\dx_{t})$ of that model. At time $0$, we simulate $X_{0}^{n}\sim \M_{0}(\dx_{0})$
for $n=1,\ldots,N$, and weight these `particles' according to function
$G_{0}$. Then 
\[
\Q_{0}^{N}(\dx_{0})=\sum_{n=1}^{N}W_{0}^{n}\delta_{X_{0}^{n}}(\dx_{0}),\qquad W_{0}^{n}=\frac{G_{0}(X_{0}^{n})}{\sum_{m=1}^{N}G_{0}(X_{0}^{m})}
\]
is an importance sampling approximation of $\Q_{0}(\dx_{0})$, in
the sense that 
\[
\Q_{0}^{N}(\varphi)=\sum_{n=1}^{N}W_{0}^{n}\varphi(X_{0}^{n})\approx\Q_{0}(\varphi)
\]
 for any suitable test function $\varphi$. 

To progress to time 1, recall from (\ref{eq:forward_recursion}) that
\[
\Q_{1}(\dx_{\text{0}:1})=\frac{1}{\ell_{1}}\Q_{0}(\dx_{0})M_{1}(x_{0},\dx_{1})G_{1}(x_{0},x_{1})
\]
which suggests to perform importance sampling, with proposal $\Q_{0}(\dx_{0})M_{1}(x_{0},\dx_{1})$,
and weight function $G_{1}$. But since $\Q_{0}(\dx_{0})$ is not
available, we use instead $\Q_{0}^{N}$: that is, we sample $N$ times
from 
\[
\sum_{n=1}^{N}W_{0}^{n}\delta_{X_{0}^{n}}(\dx_{0})M_{1}(X_{0}^{n},\dx_{1}).
\]
 To do so, for each $n$, we draw $A_{1}^{n}\sim\mathcal{M}(W_{0}^{1:N})$,
the multinomial distribution which generates value $m$ with probability
$W_{0}^{m}$; then we sample $X_{1}^{n}\sim M_{1}(X_{0}^{A_{1}^{n}},\dx_{1})$.
We obtain in this way $N$ pairs $(X_{0}^{A_{1}^{n}},X_{1}^{n})$,
and we re-weight them according to function $G_{1}$. In particular
\[
\Q_{1}^{N}(\dx_{1})=\sum_{n=1}^{N}W_{1}^{n}\delta_{X_{1}^{n}}(\dx_{1}),\qquad W_{1}^{n}=\frac{G_{1}(X_{0}^{A_{1}^{n}},X_{1}^{n})}{\sum_{m=1}^{N}G_{1}(X_{0}^{A_{1}^{m}},X_{1}^{m})}
\]
is our approximation of $\Q_{1}(\dx_{1})$. 

We proceed similarly at times 2, 3, ...; see Algorithm \ref{alg:SMC}.
At every time $t$, we sample $N$ points from 
\[
	\sum_{n=1}^N W_{t-1}^n \delta_{X_{t-1}^n} (\dx_{t-1}) M_t(X_{t-1}^n,\dx_t)
\]
and assign weights 
$W_t^n \propto G_t(X_{t-1}^{A_t^n}, X_t^n)$ to the so-obtained pairs $(X_{t-1}^{A_t^n}, X_t^n)$.  
Then we may use 
\[
	\sum_{n=1}^{N}W_{t}^{n}\varphi(X_{t}^{n}) 
\]
as an approximation of $\Q_{t}(\varphi)$, for any test function $\varphi:\setX\rightarrow\R$.
The approximation error of $\Q_t(\varphi)$ converges to zero at rate $\bigO_{P}(N^{-1/2})$,
under appropriate conditions \cite{DelGui,Chopin:CLT}.

\begin{algorithm}
Step $0$:
\begin{description}
\item [{(a)}] Sample $X_{0}^{n}\sim \M_{0}(\dx_{0})$ for $n=1,\ldots,N$.
\item [{(b)}] Compute weight $W_{0}^{n}=G_{0}(X_{0}^{n})/\sum_{m=1}^{N}G_{0}(X_{0}^{m})$
for $n=1,\ldots,N$. 
\end{description}
Recursively, for $t=1,\ldots,T$:
\begin{description}
\item [{(a)}] Sample $A_{t}^{1:N}\sim\mathcal{M}(W_{t-1}^{1:N})$; see
Appendix A. 
\item [{(b)}] Sample $X_{t}^{n}\sim M_{t}(X_{t-1}^{A_{t}^{n}},\dx_{t})$
for $n=1,\ldots,N$.
\item [{(c)}] Compute weight $W_{t}^{n}=G_{t}(X_{t-1}^{A_{t}^{n}},X_{t}^{n})/\sum_{m=1}^NG_{t}(X_{t-1}^{A_{t}^{m}},X_{t}^{m})$
for $n=1,\ldots,N$.
\item [{\caption{\label{alg:SMC}Generic SMC sampler, for a given Feynman-Kac model}
}]~
\end{description}
\end{algorithm}

\subsection{Back to State-Space Models\label{subsec:Back-to-SSMs}}

We have explained in Section \ref{subsec:Feynman-Kac-in-practice}
that, for a given state-space model, there is an infinite number
of Feynman-Kac models such that $\Q_{t}(\dx_{t})$ is the filtering distribution.
Thus, there is also an infinite number of SMC algorithms that may
be used to approximate this filtering distribution. 
\begin{example}
The Feynman-Kac model defined in Example \ref{exa:rare-event} is
such that $\Q_{t}(\dx_{t})$ is the distribution of $X_{t}$ conditional
on $X_{s}\geq0$ for all $0\leq s\leq t$, where $\left\{ X_{t}\right\} $
is a Gaussian auto-regressive process: $X_{t}=\phi X_{t-1}+V_{t}$,
$V_{t}\sim N(0,1)$. We may interpret $\Q_{t}(\dx_{t})$ as the filtering
distribution of a state-space model, where $\left\{ X_{t}\right\} $
is the same auto-regressive process, $Y_{t}=\ind_{\Rp}(X_{t})$, and
$y_{t}=1$ for all $t$. Consider now the following alternative Feynman-Kac
model: $M_{t}(x_{t-1},\dx_{t})$ is the Normal distribution $N(\phi x_{t-1},1)$
truncated to $\Rp$, i.e. the distribution with probability density 

\[
m_{t}(x_{t}|x_{t-1})=\frac{\varphi(x_{t}-\phi x_{t-1})}{\Phi(\phi x_{t-1})}\ind_{\Rp}(x_{t})
\]
where $\varphi$ and $\Phi$ are respectively the PDF and CDF of a
$N(0,1)$ distribution; and $G_{t}(x_{t-1},x_{t})=\Phi(\phi x_{t-1})$,
as per (\ref{eq:guided_Gt}). Again, quick calculations show that
we recover exactly the same distributions $\Q_{t}(\dx_{t})$. Hence
we have two SMC algorithms that approximate the same sequence of distributions
(one for each Feynman-Kac model). Observe however that the latter
SMC algorithm simulates all particles directly inside the region of
interest ($\Rp)$, while the former (bootstrap) algorithm simulates
particles `blindly', and assigns zero weight to those particles that
fall outside $\R^{+}$. As a result, the latter algorithm tends to
perform better. Note also that, under both Feynman-Kac formulations, $L_{t}$
is the probability that $X_{s}\geq0$ for all $0\leq s\leq t$, hence
both algorithms may be used to approximate this rare-event probability
(see Section \ref{subsec:Extensions} below), but again the latter
algorithm should typically give lower variance estimates for $L_{t}$. 
\end{example}
Of course, the previous example is a bit simplistic, as far as state-space
models are concerned. Recall from Section \ref{subsec:Feynman-Kac-in-practice}
that, for a given state-space model, any Feynman-Kac model such that
$G_{t}$ is set to (\ref{eq:guided_Gt}) recovers the filtering distribution
of that model for $\Q_{t}$. The usual recommendation is to choose
one such Feynman-Kac model in a way that the variance of the weights
of the corresponding SMC algorithm is low. To minimise the variance
of the weights at iteration $t$, one should take \cite{DouGodAnd}
the guided Feynman-Kac model such that 
\[
M_{t}^{\mathrm{opt}}(x_{t-1},\dx_{t})\propto P_{t}(x_{t-1},\dx_{t})f_{t}(y_{t}|x_{t}),
\]
the distribution of $X_{t}|\big(X_{t-1}=x_{t-1},Y_{t}=y_{t}\big)$. In words,
one should \emph{guide }particles to a part of space $\setX$ where
likelihood $x_{t}\rightarrow f_{t}(y_{t}|x_{t})$ is high. 

In fact, in the previous example, the second Feynman-Kac model corresponds
precisely to this optimal kernel. Unfortunately, for most models sampling
from the optimal kernel is not easy. One may instead derive an easy-to-sample
kernel $M_{t}$ that approximates $M_{t}^{\mathrm{opt}}$ in some
way. Again, provided $G_{t}$ is set to (\ref{eq:guided_Gt}), one
will recover the exact filtering distribution as $\Q_{t}$.
\begin{example}
In Example \ref{exa:StochVol}, \cite{PittShep} observed that the
bootstrap filter performs poorly at iterations $t$ where the data-point
$y_{t}$ is an outlier (i.e. takes a large absolute value). A potential
remedy is to take into account $y_{t}$ in some way when simulating
$X_{t}$. To simplify the discussion, take $\mu=0$, and consider
the probability density of $X_{t}|X_{t-1},Y_{t}$:
\begin{align*}
p_{t}(x_{t}|x_{t-1},y_{t}) & \propto\varphi( (x_{t}-\phi x_{t-1})/\sigma)\varphi(y_{t};0,e^{x_{t}})\\
 & \propto\exp\left\{ -\frac{1}{2\sigma^{2}}(x_{t}-\phi x_{t-1})^{2}-\frac{x_{t}}{2}-\frac{y_{t}^{2}}{2e^{x_{t}}}\right\} .
\end{align*}
It is not easy to simulate from this density, but \cite{PittShep}
suggested to approximate it by linearizing $\exp(-x_{t})$ around
$x_{t}=\phi x_{t-1}$: $\exp(-x_{t})\approx\exp(-\phi x_{t-1})(1+\phi x_{t-1}-x_{t})$.
This leads to proposal density 
\[
m_{t}(x_{t}|x_{t-1})\propto\exp\left\{ -\frac{1}{2\sigma^{2}}(x_{t}-\phi x_{t-1})^{2}-\frac{x_{t}}{2}-\frac{y_{t}^{2}}{2e^{\phi x_{t-1}}}(1+\phi x_{t-1}-x_{t})\right\} 
\]
which is clearly Gaussian (and hence easy to simulate from). Note
that this linear `approximation' does not imply that the resulting
SMC algorithm is approximate in some way: provided $G_{t}$ is set
to (\ref{eq:guided_Gt}), the resulting algorithm targets exactly
the filtering distribution of the model, as we have already discussed.
\end{example}

\subsection{Sequential quasi-Monte Carlo }

\subsubsection{QMC Basics}

As mentioned in the introduction, we assume that the reader is already
familiar with QMC and RQMC (randomised QMC); otherwise see e.g. the
books of \cite{Lemieux:MCandQMCSampling} and \cite{MR3289176}.
We only recall briefly the gist of QMC. Consider an expectation with
respect to $\Unif\left([0,1]^{d}\right)$, and its standard Monte
Carlo approximation: 
\[
\frac{1}{N}\sum_{n=1}^{N}\varphi(U^{n})\approx\int_{[0,1]^{d}}\varphi(u)\,\dd u
\]
where the $U^{n}$ are IID variables. QMC amounts to replacing the
$U^{n}$ by $N$ deterministic points $u^{n,N}$ that have low discrepancy.
The resulting error converges faster than with Monte Carlo under certain
conditions, in particular regarding the \emph{regularity }of function
$\varphi$. This is an important point when it comes to apply QMC
in practice: rewriting a given algorithm as a deterministic function
of uniforms, and replacing these uniforms by a QMC point set, may
not warrant better performance. One has also to make sure that this
deterministic function is indeed regular, and maintain low discrepancy
in some sense. 

\subsubsection{SQMC when $d=1$\label{subsec:SQMC-dim-one}}

We explained in Section \ref{subsec:SMC} that SMC amounts to a sequence
of importance sampling steps, with proposal distribution 

\begin{equation}
\sum_{n=1}^{N}W_{t-1}^{n}\delta_{X_{t-1}^{n}}(\dx_{t-1})M_{t}(x_{t-1},\dx_{t})\label{eq:mixture}
\end{equation}
at time $t$. To derive a QMC version of this algorithm, we must find
a way to generate a low-discrepancy sequence with respect to this
distribution. The difficulty lies in the  fact that the support
of (\ref{eq:mixture}) is partly discrete (the choice of the ancestor
$X_{t-1}^{n})$, partly continuous (the kernel $M_{t}(x_{t-1},\dx_{t})$).
We focus on the discrete part below. For the continuous part, we assume
that $\setX\subset\R^{d}$, and that we know of a function $\Gamma_{t}:\setX\times[0,1]^{d}\rightarrow\setX$ 
such that, for any $x_{t-1}\in\setX$, $\Gamma_{t}(x_{t-1},U)$, $U\sim\mathcal{U}\left([0,1]^{d}\right)$,
has the same distribution as $M_{t}(x_{t-1},\dx_{t})$. The choice
of $\Gamma_{t}$ is model-dependent, and is often easy; the default
choice would be the Rosenblatt transform associated to $M_{t}(x_{t-1},\dx_{t})$
(the multivariate inverse CDF).
\begin{example}
Consider a state-space model with latent process $X_{t}=\phi X_{t-1}+V_{t}$,
$V_{t}\sim N(0,\sigma^{2})$. Then one would take typically $\Gamma_{t}(x_{t-1},u)=\phi x_{t-1}+\sigma\Phi^{-1}(u)$,
where $\Phi$ is the CDF of a $N(0,1)$ distribution. In dimension
$d>1$, such a process would take the form $X_{t}=AX_{t-1}+V_{t}$,
$V_{t}\sim N(0,\Sigma)$, where $A$ is a $d\times d$ matrix. Then
one would define $\Gamma_{t}(x_{t-1},u)=Ax_{t-1}+\Pi_{\Sigma}(u)$,
where the second term may be defined in several ways; e.g. (a) $\Pi_{\Sigma}(u)$
is the Rosenblatt transform of $N(0,\Sigma),$ i.e. first component
of $\Pi_{\Sigma}(u)$ is $\Sigma_{11}^{1/2}\Phi^{-1}(u_{1})$ and
so on; or (b) $\Pi_{\Sigma}(u)=C\boldsymbol{\Phi}^{-1}(u)$, where
$C$ is the Cholesky lower triangle of $\Sigma$, $CC^{T}=\Sigma$,
and $\boldsymbol{\Phi}^{-1}$ is the function which assigns to vector
$u$ the vector $\left(\Phi^{-1}(u(1)),\ldots,\Phi^{-1}(u(d))\right)^{T}$.
In both cases, function $\Gamma_{t}$ depends on the order of the
components of $X_{t}$. 
\end{example}

We now focus on the discrete component of (\ref{eq:mixture}). The
standard approach to sample from such a finite distribution is the
inverse CDF method: define $F_{t-1}^{N}(x)=\sum_{n=1}^{N}W_{t-1}^{n}\ind\left\{ n\leq x\right\} $,
and set $\hat{X}_{t-1}^{n}=X_{t-1}^{A_{t}^{n}}$ with $A_{t}^{n}=\left(F_{t-1}^{N}\right)^{-1}(U_{t}^{n})$,
where $U_{t}^{n}\sim\Unif\left([0,1]\right)$ and $\left(F_{t}^{N}\right)^{-1}$
is the generalised inverse of $F_{t}^{N}$. This is precisely how
resampling is implemented in a standard particle filter. See Appendix
A for a description of the standard algorithm to evaluate in $\bigO(N)$
time function $\left(F_{t-1}^{N}\right)^{-1}$ for $N$ inputs. 

A first attempt at introducing a QMC point set would be to set again
$A_{t}^{n}=\left(F_{t}^{N}\right)^{-1}(U_{t}^{n}),$ but taking this
time for $U_{t}^{n}$ the first component of a QMC point set (of dimension
$d+1$). The problem with this approach is that this defines a transformation,
from the initial uniforms to the points, which is quite irregular.
In fact, since the labels of the $N$ particles are arbitrary, this
distribution somehow involves a \emph{random permutation} of the $N$
initial points. In other terms, we add some noise in our transformation,
which is not a good idea in  any type of QMC procedure. 

Now consider the special case $\setX\subset\R$, and let $\sigma_{t-1}=\mathrm{argsort}(X_{t-1}^{1:N})$,
i.e. $\sigma_{t-1}$ is a permutation of the $N$ first integers such
that: 
\[
X_{t-1}^{\sigma_{t-1}(1)}\leq\ldots\leq X_{t-1}^{\sigma_{t-1}(N)}
\]
and, for $x\in\setX$, let 
\[
\hat{F}_{t-1}^{N}(x)=\sum_{n=1}^{N}W_{t-1}^{n}\ind\left\{ X_{t-1}^{n}\leq x\right\} =\sum_{n=1}^{N}W_{t-1}^{\sigma_{t-1}(n)}\ind\left\{ X_{t-1}^{\sigma_{t-1}(n)}\leq x\right\} .
\]
Note that $\hat{F}_{t-1}^{N}$ does not
depend on the \emph{labels} of the $N$ ancestors (like $F_{t-1}^{N}$ does);
for instance, the smallest $x$ such that $\hat{F}_{t-1}^{N}(x)>0$
is $X_{t-1}^{\sigma_{t-1}(1)}$, the smallest ancestor (whatever its label). 

The first main idea in SQMC is to choose $A_{t}^{n}$ such that $X_{t-1}^{A_{t}^{n}}=\hat{F}_{t-1}^{N}(U_{t}^{n})$,
where $U_{t}^{n}$ is the first component of some QMC or RQMC point
set. In this way, the resampled ancestors, i.e. the points $X_{t-1}^{A_{t}^{n}}$,
may be viewed as a low-discrepancy point set with respect to the marginal
distribution of component $x_{t-1}$ in distribution (\ref{eq:mixture}).
In practice, computing $A_{t}^{n}$ amounts
to (a) sort the $N$ ancestors; and (b) apply the inverse CDF algorithm
of Appendix A to these $N$ sorted ancestors. 

\subsubsection{SQMC for $d>1$ }

When $\setX\subset\R^{d}$, with $d>1$, it is less clear how to invert
the empirical CDF of the ancestors 
\[
\hat{F}_{t-1}^{N}(x)=\sum_{n=1}^{N}W_{t-1}^{n}\ind\left\{ X_{t}^{n}\leq x\right\} 
\]
as this function is $\R^{d}\rightarrow[0,1]$. 

The second main idea in SQMC is to transform the $N$ ancestors $X_{t-1}^{n}$
into $N$ scalars $Z_{t-1}^{n}$, in a certain way that maintains
the low discrepancy of the $N$ initial points. Then we may construct
a QMC point relative to 
\[
\hat{F}_{t-1,h}^{N}(z)=\sum_{n=1}^{N}W_{t-1}^{n}\delta_{Z_{t-1}^{n}}(\dd z),\quad z\in[0,1]
\]
in the same way as described in the previous section. 

To do so, we take $Z_{t-1}^{n}=h\circ\psi(X_{t-1}^{n})$, where
$h:[0,1]^{d}\rightarrow[0,1]$ is the inverse of the Hilbert curve,
see below, and $\psi:\setX\rightarrow[0,1]^{d}$ is model-dependent.
(For instance, if $\setX=\R^{d}$, we may apply a component-wise
version of the logistic transform.)

\begin{figure}
\centering
\includegraphics[scale=0.1]{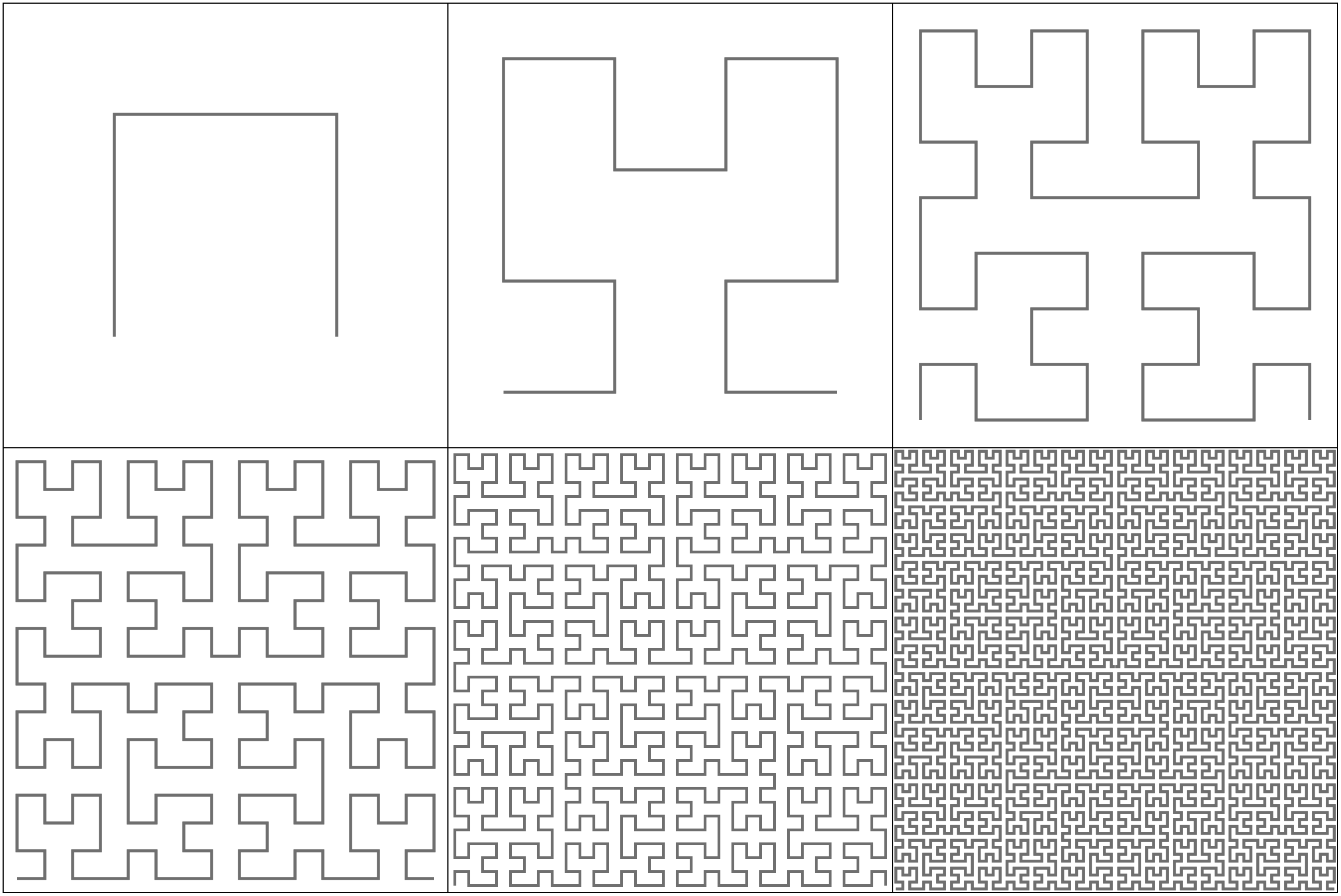} 
\caption{\label{fig:Hilbertcurve}Sequence of curves of which the Hilbert curve
is the limit, for $d=2$ (Source: Wikipedia)}
\end{figure}

The Hilbert curve is a space-filling curve, that is a function $H:[0,1]\rightarrow[0,1]^{d}$
with the following properties: it is defined as the limit of the process
depicted in Figure \ref{fig:Hilbertcurve}; it is H\"older with coefficient
$1/d$ (in particular it is continuous); it `fills' entirely $[0,1]^{d}$;
the set of points in $[0,1]^{d}$ that admit more than one pre-image
is of measure $0$. Thanks to these properties, it is possible to
define a pseudo-inverse $h:[0,1]\rightarrow[0,1]^{d}$, such that
$H\circ h(u)=u$ for $u\in[0,1]$.

In addition, the pseudo-inverse $h$ maintains low-discrepancy in the
following sense: if the $N$ ancestors $X_{t-1}^{n}$ are such that
$\|\pi^{N}-\pi\|_{E}\rightarrow0$ where $\pi^{N}(\dx)=\sum_{n=1}^{N}W_{t-1}^{n}\delta_{X_{t-1}^{n}}(\dx)$,
and $\pi$ is some limiting probability distribution, then (under
appropriate conditions, see Theorem 3 in \cite{MR3351446}), $\|\pi_{h}^{N}-\pi_{h}\|_{E}\rightarrow0$,
where $\pi_{h}^{N}$ and $\pi_{h}$ are the images of $\pi^{N}$ and
$\pi$ through $h$. The extreme norm $\|\cdot\|_{E}$ in this theorem
is some generalisation of the QMC concept of extreme discrepancy;
again see \cite{MR3351446} for more details.

We note that other functions $[0,1]^{d}\rightarrow[0,1]$ (e.g. pseudo-inverse
of other space-filling curves, such as the Lebesgue curve) could be used
in lieu of the inverse of the Hilbert curve. However, our impression
is that other choices would not necessarily share the same property
of ``maintaining low discrepancy''. At the very least, our proofs
in \cite{MR3351446} rely on properties that are specific to the
Hilbert curve, and would not be easily extended to other functions. 

Algorithm \ref{alg:SQMC} summarises the operations performed in SQMC. 

\begin{algorithm}
At time $0$, 
\begin{description}
\item [{(a)}] Generate a QMC point set $u_{0}^{1:N}$ of dimension $d$.
\item [{(b)}] Compute $X_{0}^{n}=\Gamma_{0}(u_{0}^{n})$ for all $n\in1:N$. 
\item [{(b)}] Compute $W_{0}^{n}=G_{0}(X_{0}^{n})/\sum_{m=1}^{N}G_{0}(X_{0}^{m})$
for all $n\in1:N$. 
\end{description}
Recursively, for time $t=1:T,$ 
\begin{description}
\item [{(a)}] Generate a QMC or RQMC point set $(u_{t}^{1:N},v_{t}^{1:N})$
of dimension $d+1$ ($u_{t}^{n}$ being the first component, and
$v_{t}^{n}$ the vector of the $d$ remaining components, of point
$n$). 
\item [{(b)}] Hilbert sort: find permutation $\sigma_{t}$ such that $h\circ\psi(X_{t-1}^{\sigma_{t}(1)})\leq\ldots\leq h\circ\psi(X_{t-1}^{\sigma_{t}(N)})$
if $d\geq2$, or $X_{t-1}^{\sigma(1)}\leq\ldots\leq X_{t-1}^{\sigma(N)}$
if $d=1$. 
\item [{(c)}] Generate $A_{t}^{1:N}$ using Algorithm \ref{alg:inverse_method},
with inputs $\mathrm{sort}(u_{t}^{1:N})$ and $W_{t}^{\sigma(1:N)}$,
and compute $X_{t}^{n}=\Gamma_{t}(X_{t-1}^{\sigma_{t}(A_{t}^{n})},v_{t}^{n})$. 
\item [{(e)}] Compute $W_{t}^{n}=G_{t}(X_{t-1}^{\sigma(A_{t}^{n})},X_{t}^{n})/\sum_{m=1}^{N}G_{t}(X_{t-1}^{\sigma(A_{t}^{m})},X_{t}^{m})$
for all $n\in1:N$. 
\end{description}
\caption{\label{alg:SQMC} SQMC algorithm}
\end{algorithm}

\subsection{Connection to Array-RQMC}

In the Feynman-Kac formalism, taking $G_{0}(x_{0})=1,$ $G_{t}(x_{t-1},x_{t})=1$
for all $t\geq1$, makes $\Q_{t}$ the distribution of the Markov
chain $\{X_{t}\}$. In that case, SQMC may be used to approximate
expectations with respect to the distribution of that Markov chain.
In fact, such a SQMC algorithm may be seen as a certain version of
the array-RQMC algorithm of \cite{LEcuyer2006}, where the particles
are ordered at every iteration using the inverse of the Hilbert curve.
In return, the convergence results established in \cite{MR3351446}
apply to that particular version of array-RQMC. 

Although designed initially for a smaller class of problems, array-RQMC
is built on the same insight as SQMC of viewing the problem of interest
not a single Monte Carlo exercise, of dimension $d(T+1)$ (e.g. simulating
a Markov chain in $\setX\subset R^{d}$ over $T+1$ time steps), but
as $T+1$ exercises of dimension $d+1$. See also \cite{Fearnhead2005}
for a related idea in the filtering literature. 

\subsection{Extensions\label{subsec:Extensions}}

In state-space modelling, one may be interested in computing other
quantities than the filtering distributions: in particular the likelihood
of the data up to $t$, $p_{t}(y_{0:t})$, and the smoothing distribution,
i.e. the joint law of the states $X_{0:T}$, given some complete dataset
$Y_{0:T}$. 

The likelihood of the data $p_{t}(y_{0:t})$ equals the normalising
constant $L_{t}$ in any guided Feynman-Kac model. This quantity may
be estimated at iteration $t$ as follows: 

\[
L_{t}^{N}=\left(\frac{1}{N}\sum_{n=1}^{N}G_{0}(X_{0}^{n})\right)\prod_{s=1}^{t}\left(\frac{1}{N}\sum_{n=1}^{N}G_{s}(X_{s-1}^{A_{s}^{n}},X_{s}^{n})\right).
\]
A non-trivial property of SMC algorithms is that this quantity is
an unbiased estimate of $L_{t}$ \cite{DelMoral1996unbiased}. This
makes it possible to develop MCMC algorithms for parameter estimation
of state-space models which (a) runs at each MCMC iteration a particle
filter to approximate the likelihood at given value of the parameter;
and yet (b) targets the exact posterior distribution of the parameters,
despite the fact the likelihood is computed only approximately. The
corresponding PMCMC (particle MCMC) algorithms have been proposed
in the influential paper of \cite{PMCMC}. If we use RQMC (randomised
QMC) point steps within SQMC, then $L_{t}^{N}$ remains an unbiased
estimate of $L_{t}$. Thus, SQMC is compatible with PMCMC (meaning
that one may use SQMC instead of SMC at every iteration of a PMCMC
algorithm), and in fact one may improve the performance of PMCMC in
this way; see \cite{MR3351446} for more details. 

Smoothing is significantly more difficult than filtering. Smoothing
algorithms usually amount to (a) run a standard particle filter,
forward in time; (b) run a second algorithm, which performs some operations
on the output of the first algorithm, backward in time. Such algorithms
have complexity $O(N^{2})$ in general. We refer the readers to \cite{MR2577439},
\cite{Doucet2009} for a general presentation of smoothing algorithms,
and to \cite{arxiv:1506.06117} for how to derive QMC smoothing algorithms
that offer better performance than standard (Monte Carlo-based) smoothing
algorithms. 

Finally, we mention that SMC algorithms may also be used in other
contexts that the sequential inference of state-space models. Say
we wish to approximate expectations with respect to some distribution
of interest $\pi$, but it is is difficult to sample directly from
$\pi$ (e.g. the density $\pi$ is strongly multimodal). One may define
a geometric bridge between some easy to sample distribution $\pi_{0}$
and $\pi$ as follows: $\pi_{t}(x)\propto\pi_{0}(x)^{1-\gamma_{t}}\pi(x)^{\gamma_{t}}$
where $0=\gamma_{0}<\ldots<\gamma_{T}=1$. Then one may apply SMC
to the sequence $\left(\pi_{t}\right)$, and use the output of the
final iteration to approximate $\pi$. Other sequence of distributions
may be considered as well. For more background on such applications
of SMC see e.g. \cite{MR1837132}, \cite{MR1929161}, and \cite{DelDouJas:SMC}.
The usefulness of SQMC for such problems remains to be explored. 

\subsection{A Note on the Impact of the Dimension }

\cite{MR3351446} include a numerical study of the impact of the
dimension on the performance of SQMC. It is observed that the extra
performance of SQMC (relative to standard SMC) quickly decreases with
the dimension. 

Three factors may explain this curse of dimensionality: 
\begin{enumerate}
\item The inherent curse of dimensionality of QMC: 
the standard discrepancy bounds invoked as a formal justification of QMC 
deteriorate with the dimension. 
\item Regularity of the Hilbert curve: the Hilbert curve is H\"older with
	coefficient $1/d$. Consequently, the mapping  $u_t^n \mapsto
	X_{t-1}^{\sigma_t(A_{t-1}^n)}$ induced by steps (a) and (b) of
	Algorithm \ref{alg:SQMC} for time $t\geq 1$ is less and less regular as
	the dimension increases. (We however believe that this property is not
	specific to the use of the Hilbert curve but is due to  the resampling
	mechanism itself, where a single point in $u_t^n\in [0,1]$ is used to
	select the $d$-dimensional ancestor $X_{t-1}^{A_t^n}$.)
%

\item SMC curse of dimensionality: SMC methods also suffer from the curse
of dimensionality, for the simple reason that they rely on importance
sampling: the larger the dimension, the greater the discrepancy between
the proposal distribution and the target distribution. In practice,
one observes in high-dimensional filtering problem that, at each iteration,
only a small proportion of the particles get a non-negligible weight. 
\end{enumerate}
We thought earlier that factor 2 was the `main culprit'. However,
factor 3 seems to play an important part as well. To see this, we
compare below the relative performance of SQMC and SMC for the filtering
of the following class of linear Gaussian state-space models (as in
\cite{2015arXiv151106286G}): $X_{0}\sim N_{d}(0,I_{d})$, and 
\begin{align*}
X_{t} & =FX_{t-1}+V_{t},\qquad V_{t}\sim N_{d}(0,I_{d}),\\
Y_{t} & =X_{t}+W_{t},\qquad W_{t}\sim N_{d}(0,I_{d}),
\end{align*}
with $F=(\alpha^{\left|i-j\right|})_{i,j=1:d}$, and $\alpha=0.4$.
For such models, the filtering distribution may be computed exactly
using  the Kalman filter \cite{KalBuc}. We consider two Feynman-Kac
formalisms of that problem:
\begin{itemize}
\item The bootstrap formalism, where $M_{t}$ is set to $N_{d}(FX_{t-1},I_{d})$,
the distribution of $X_{t}|X_{t-1}$ according to the model, and $G_{t}(x_{t-1},x_{t})=f_{t}(y_{t}|x_{t})=N_{d}(y_{t};x_{t},I_{d})$,
the probability density at point $y_{t}$ of distribution $N_{d}(x_{t-1},I_{d})$.
\item The `optimal' guided formalism where 
\[
M_{t}(x_{t-1},\dx_{t})\propto P_{t}(x_{t-1},\dx_{t})f_{t}(y_{t}|x_{t})\sim N_{d}\left(\frac{Y_{t}+FX_{t-1}}{2},\frac{1}{2}I_{d}\right)
\]
and, by (\ref{eq:guided_Gt}), 
\[
G_{t}(x_{t-1},x_{t})=N_{d}(y_{t};Fx_{t-1},2I_{d})
\]
the probability density at point $y_{t}$ of distribution $N_{d}(Fx_{t-1},2I_{d})$.
\end{itemize}
In both cases, as already explained, we recover the filtering distribution
as $\Q_{t}$. But the latter formalism is chosen so as to minimise
the variance of the weights at each iteration.

\begin{figure}[h]
	\centering
	\includegraphics[scale=0.5]{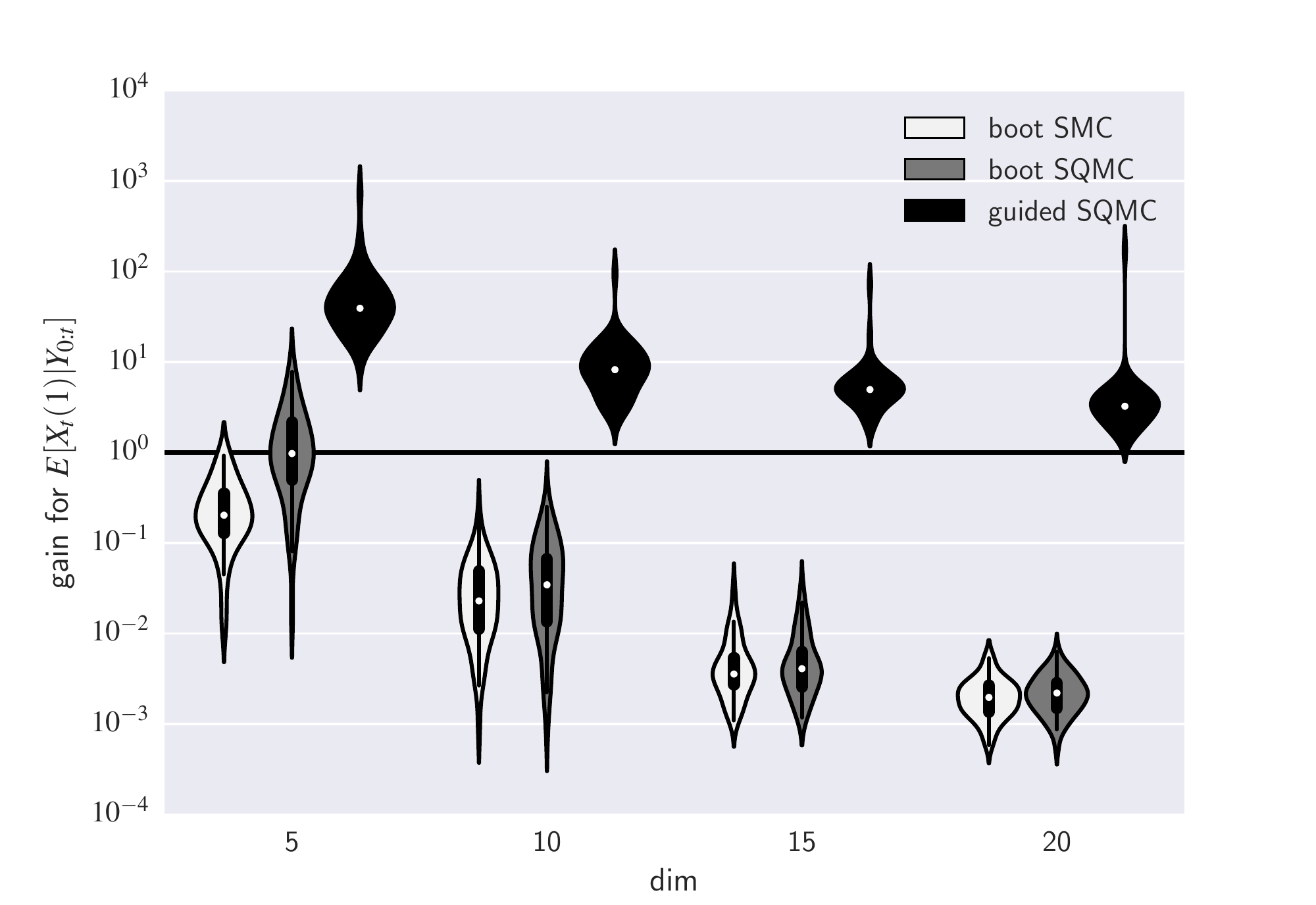}
	\caption{Violin plots of the gains of the considered algorithms 
		when estimating the filtering expectations $\mathbb{E}[X_t(1)|Y_{0:t}]$
		for $t=0,\ldots,T=50$. (Each violin plot represents the variability 
		of the $T$ gains for these $T$ estimates.) 
		Gain is MSE (mean square error) of reference algorithm (guided-SMC) divided 
		by MSE of considered algorithm.
		\label{fig:violins}
	}
\end{figure}

We simulate $T=50$ data-points from the model, for $d=5$, 10, 15 and 20.
Figure \ref{fig:violins} compares the following four algorithms: 
SMC-bootstrap, SQMC-bootstrap, SMC-guided, and SQMC-guided. The comparison is 
in terms of the MSE (mean square error) of the estimate of 
the filtering expectation of the first component of $X_t$, i.e. 
$\mathbb{E}[X_t(1)|Y_{0:t}=y_{0:t}]$. We use SMC-guided as the reference algorithm, and we plot 
for each of the three other algorithms the variations of the gain (MSE of reference 
algorithm divided by MSE of considered algorithm) for the $T$ estimates. 
(We use violin plots, which are similar to box-plots, except that the box is replaced 
by kernel density estimates.) 
A gain $g$ means that the considered 
algorithm would need $g$ times less particles (roughly) to provide an estimate with 
a similar variance (to that of the reference algorithm). Each algorithm was run 
with $N=10^4$. 

First, we observe that guided algorithms outperforms bootstrap algorithms more and more 
as the dimension increases. Second, for bootstrap algorithms, the performance between 
SMC and SQMC is on par as soon as $d\geq 10$. (In fact, the performance is rather bad 
in both cases, owning to the aforementioned curse of dimensionality.) 
On the other hand, for guided formalisms we still observe a gain of order
$\bigO(10^1)$ (resp. $10^{0.5}$) for $d=10$ (resp. $d=20$).

The bottom line is that the amount of extra performance brought by SQMC
(relative to SMC) depends strongly on the chosen Feynman-Kac formalism. 
If one is able to construct a Feynman-Kac formalism 
(for the considered problem) that leads to good performance for the corresponding 
SMC algorithm (meaning that the variance of the weights is low at each iteration),
then one may expect significant extra performance from SQMC, even in high dimension.

\section{Application to Diffusions}

\subsection{Dimension Reduction in SQMC}

We start this section by a basic remark, which makes it possible to
improve the performance of SQMC when applied to models having a certain
structure. We explained in Section \ref{subsec:SMC} that SMC amounts
to performing importance sampling at every step, using as a proposal
distribution: 
\begin{equation}
\sum_{n=1}^{N}W_{t-1}^{n}\delta_{X_{t-1}^{n}}(\dx_{t-1})M_{t}(x_{t-1},\dx_{t})\label{eq:mixture_again}
\end{equation}
and as a target distribution, the same distribution times $G_{t}(x_{t-1},x_{t})$
(up to a constant). We used this remark to derive SQMC as an algorithm
that constructs a low-discrepancy point-set with respect to the distribution
above; i.e. to construct $N$ points $\big(X_{t-1}^{A_{t}^{n}},X_{t}^{n}\big)$,
the empirical distribution of which approximates well (\ref{eq:mixture_again}).

Now consider a situation where we know of a function $\Lambda:\setX\rightarrow\R^{k}$,
with $k<d$, such that (a) $G_{t}$ depends only on $X_{t}$ and $\Lambda(X_{t-1})$;
and Markov kernel $M_{t}(x_{t-1},\dx_{t})$ also depends only on $\Lambda(x_{t-1})$.
(In particular, it is possible to simulate $X_{t}$ conditional on
$X_{t-1}$, knowing only $\Lambda(X_{t-1})$.) In that case, one may
define the same importance sampling operation on a lower-dimensional
space. In particular, the new proposal distribution would be: 

\[
\sum_{n=1}^{N}W_{t-1}^{n}\delta_{\Lambda(X_{t-1}^{n})}(\dd\lambda_{t-1})M_{t}^{\Lambda}(\lambda_{t-1},\dx_{t})
\]
where $M_{t}^{\Lambda}(\lambda_{t-1},\dx_{t})$ is simply the Markov
kernel which associates distribution $M_{t}(x_{t-1},\dx_{t})$ to
any $x_{t-1}$ such that $\Lambda(x_{t-1})=\lambda_{t-1}$. We may use exactly
the same ideas as before, i.e. generate a QMC point of dimension $d+1$,
and use the first component to pick the ancestor. However, the Hilbert
sorting is now applied to the $N$ points $\Lambda(X_{t-1}^{n})$,
and therefore operates in a smaller dimension. Thus one may expect
better performance, compared to the standard version of SQMC.

This remark is related somehow to the QMC notion of ``effective dimension'':
the performance of QMC may remain good in high-dimensional problems,
if one is able to reformulate the problem in such a way that it depends 
``mostly'' (or in our case, ``only'') on a few dimensions of the
state-space.

\subsection{Filtering of Diffusion Processes\label{sub:diffusion} }

We now consider the general class of diffusion-driven state-space
models: 
\begin{align*}
\dd\widetilde{X}_{t} & =\mu_{X}(\widetilde{X}_{t})+\sigma_{X}(\widetilde{X}_{t})\dd W_{t}^{X}\\
\dd\widetilde{Y}_{t} & =\mu_{Y}(\widetilde{X}_{t})+\sigma_{Y}(\widetilde{X}_{t})\dd W_{t}^{Y}
\end{align*}
where ($W_{t}^{X})_{t\geq0}$ and ($W_{t}^{Y})_{t\geq0}$ are possibly
correlated Wiener processes. Functions $\mu_{X}$, $\mu_{Y}$, $\sigma_{X}$
and $\sigma_{Y}$ may also depend on $t$, and $\mu_{Y}$, $\sigma_{Y}$
may also depend on $Y_{t}$, but for the sake of exposition we stick
to the simple notations above. 

Filtering in continuous time amounts to recover the distribution of
$\widetilde{X}_{t}$ conditional on trajectory $y_{[0:t]}$ (i.e.
the observation of process $\{ \widetilde{Y}_t\} $ over interval
$[0,t]$). However, in most practical situations, one does not observe
process $\{ \widetilde{Y}_{t}\} $ continuously, but on a grid. To simplify, we
assume henceforth that process $\{ \widetilde{Y}_{t}\} $ is observed at times
$t\in\mathbb{N}$ and we rewrite  the above model as
\begin{align}
\dd\widetilde{X}_{t} & =\mu_{X}(\widetilde{X}_{t})+\sigma_{X}(\widetilde{X}_{t})\dd W_{t}^{X}\notag\\
\widetilde{Y}_{t+1}&=\widetilde{Y}_{t}+\int_t^{t+1}\mu_Y(\widetilde{X}_s)\dd s+\int_t^{t+1}\sigma_{Y}(\widetilde{X}_{s})\dd W_{s}^{Y}\label{eq:Y_process}.
\end{align}

It is typically too difficult to work directly in continuous
time. Thus,
as standardly done when dealing with such processes, we replace the
initial process $(\widetilde{X}_t)$ by its (Euler-) discretized version
$\{X_t\}$, with discretisation
step $\delta=1/M$, $M\geq1$.
That is, $\left\{ X_{t}\right\} $
is a $\R^{M}-$valued process, where $X_t$ is a $M$-dimensional vector representing 
the original process at times $t$, $t+1/M$, ..., $t+1-1/M$, which is defined as: 
\begin{equation}\label{eq:diff_X}
\begin{split}
X_{t}(1) & =X_{t-1}(M)+\delta\mu_{X}\left(X_{t-1}(M)\right)+\sigma_{X}(X_{t-1}(M))\left\{ W_{t+\delta}^{X}-W_{t}^{X}\right\} \\
 & \vdots\\
X_{t}(M) & =X_{t}(M-1)+\delta\mu_{X}\left(X_{t}(M-1)\right)+\sigma_{X}(X_{t}(M-1))\left\{ W_{t+1}^{X}-W_{t+1-\delta}^{X}\right\}
\end{split}
\end{equation}
and the resulting dicretization of \eqref{eq:Y_process} is given by
\begin{align}\label{eq:diff_Y}
Y_{t+1}=Y_t+\delta\sum_{m=1}^M\mu_{Y}\left(X_{t}(m)\right)+\sum_{m=1}^M\sigma_{Y}\left(X_{t}(m)\right)\big\{ W_{t+\delta m}^{Y}-W_{t+\delta(m-1)}^{Y}\big\} .
\end{align}

SQMC may be applied straightforwardly to the filtering of the discretized
model defined by \eqref{eq:diff_X} and \eqref{eq:diff_Y}. However, the choice of the $\delta=1/M$ becomes problematic.
We would like to take $M$ large, to reduce the discretization bias.
But $M$ is also the dimension of the state-space, so a large $M$
may mean a degradation of performance for SQMC (relative to SMC). 

Fortunately, the dimension reduction trick of the previous section
applies here. For simplicity, consider the bootstrap Feynman-Kac formalism
of this particular state-space model:
\begin{itemize}
\item $M_{t}(x_{t-1},\dx_{t})$ is the distribution of $X_{t}|X_{t-1}$
defined by \eqref{eq:diff_X}; observe that it only depends on $X_{t-1}(M)$,
the last component of $X_{t-1}$; 
\item $G_{t}(x_{t-1},x_{t})$ is the probability density of datapoint $y_{t}$
given $X_{t}=x_{t}$ and $Y_{t-1}=y_{t-1}$, induced by \eqref{eq:diff_X}-\eqref{eq:diff_Y}; observe that it does not depend on $x_{t-1}$ when $(W^X_t)$ and $(W^Y_t)$ are uncorrelated and that it depends on $x_{t-1}$ only through $x_{t-1}(M)$ when these two processes are correlated (see the next subsection).
\end{itemize}
Hence we may define $\Lambda(x_{t-1})=x_{t-1}(M)\in\R$. The Hilbert
ordering step may be applied to the values $Z_{t}^{n}=X_{t-1}^{n}(M)$.
In fact, since these values are scalars, there is no need to implement
any Hilbert ordering, a standard sorting is enough. 

\subsection{QMC and Brownian Motion}

We now briefly discuss how to choose $\Gamma_{t}$, the deterministic
function such that $\Gamma_{t}(x_{t-1},v)$, for $x_{t-1}\in\setX$
and $v\in[0,1]^{d}$, returns a variate from kernel $M_{t}(x_{t-1},\dx_{t})$. 

The distribution of $X_{t}|X_{t-1}$ defined in the previous section
is a simple linear transform of the distribution of a Brownian path
on a regular grid. Thus, defining function $\Gamma_{t}$ amounts to
constructing a certain function $[0,1]^{M}\rightarrow\R^{M}$ that
transforms $\mathcal{U}\left([0,1]^{M}\right)$ into the joint distribution
of $(W^X_{t+\delta},\ldots,W^X_{t+1})$, conditional on $W^X_{t}$.

It is well known in the QMC literature (e.g. Section 8.2 of \cite{MR3289176})
that there is more than one way to write the simulation of a Brownian
path as a function of uniforms, and that the most obvious way may
perform poorly when applied in conjunction with QMC. More precisely,
consider the following two approaches: 
\begin{enumerate}
\item Forward construction: simulate independently the increments $W^X_{t+\delta m}-W^X_{t+\delta(m-1)}$
from a $N(0,\delta)$ distribution.
\item Brownian bridge construction \cite{Caflisch1997}: Simulate $(W^X_{t+\delta},\dots, W^X_{t+1})$ given $W^X_t$   sequentially
 according to the Van der Corput sequence: $W_{t+\delta\lceil M/2\rceil}^{X}$,
$W_{t+\delta\lceil M/4\rceil}^{X}$, $W_{t+\delta\lceil3M/4\rceil}^{X}$
until all the components of vector $(W^X_{t+\delta},\ldots,W^X_{t+1})$
are simulated. For instance,  for $s<t'<u$, we use 
\[
W_{t'}^{X}|W_{s}^{X},W_{u}^{X}\sim N_{1}\left(\frac{u-t'}{u-s}W_{s}^{X}+\frac{t'-s}{u-s}W_{u}^{X},\frac{(u-t')(t'-s)}{u-s}\right)
\]
and the fact  that $(W_{t}^{X})$ is a Markov process (i.e. $W_{t}^{X}|W_{s}^{X}$
does not depend on $W_{s'}^{X}$ for $s'<s$). 
\end{enumerate}
In both cases, it is easy to write the simulation of $(W^X_{t+\delta},\ldots,W^X_{t+1})$
as a function of $M$ uniform variates. However, in the first case,
the obtained function depends in the same way on each of the $M$
variates, while in the second case, the function depends less and
less on the successive components. This mitigates the inherent curse
of dimensionality of QMC \cite{Caflisch1997}. 

We shall observe the same phenomenon applies to SQMC; even so for
a moderate value of $M$, interestingly. We also mention briefly the
PCA (principal components analysis) construction as another interesting
way to construct Brownian paths, and refer again to Section 8.2 \cite{MR3289176}
for a more in-depth discussion of QMC and Brownian paths. 

Lastly, although we focus on univariate diffusion processes in this section for the sake of simplicity, the above considerations also hold  for multivariate models. Notably, the Brownian bridge construction is easily  generalizable to the case where $(W^X_t)$ is a $d$-dimensional vector of correlated Wiener processes. The dimension of the QMC point set used as input of SQMC is then of size $dM+1$ and the Hilbert ordering would operate on a $d$-dimensional  space. 

\subsection{Numerical Experiments}

To illustrate the discussion of the previous subsections  we consider the following diffusion driven stochastic volatility model (e.g.\cite{Chib2004})
\begin{align*}
\dd \widetilde{X}_t&=\Big\{\kappa(\mu^X-e^{\widetilde{X}_t})e^{-\widetilde{X}_t}-0.5\omega^2e^{-X_t}\Big\}\dd t+\omega  e^{-\widetilde{X}_t/2}\dd W^X_t\\
 \widetilde{Y}_{t+1}&= \widetilde{Y}_{t}+\int_t^{t+1}\big\{\mu^Y+\beta e^{\widetilde{X}_z}\big\}\dd z+\int_t^{t+1}e^{\widetilde{X}_s/2}\dd W_s^Y 
\end{align*}
where  $(W^X_t)$ and $(W^Y_t)$ are  Wiener processes with correlation coefficient  $\rho\in (-1,1)$, $\omega>0$, $\kappa>0$ while the other parameters $\mu^Y$, $\beta$ are in $\R$.

To fit this  model into the bootstrap Feynman-Kac formalism that we consider in this section, note that, for $t\geq 0$,
\begin{align*}
\widetilde{Y}_{t+1}|\widetilde{Y}_{t}, \widetilde{X}_{[t,t+1]}\sim N\Big(\widetilde{Y}_{t}+\mu^Y+\beta\sigma_{t+1}^2+\rho Z_{t+1},\,\, (1-\rho^2) \sigma_{t+1}^2\Big)
\end{align*}
with $\sigma_{t+1}^2=\int_{t}^{t+1} e^{\widetilde{X}_s}\dd s$ and  $Z_{t+1}=\int_{t}^{t+1} e^{\widetilde{X}_s/2}\dd W^X_s$, and thus, as explained in  Section \ref{sub:diffusion},
\begin{equation*}
\begin{split}
G_t(x_{t-1},x_t)&=\tilde{G}_t(x_{t-1}(M),x_t)\\
&:=N\Big(\widetilde{Y}_{t+1}; \widetilde{Y}_{t}+\mu^Y+\beta \hat{\sigma}_{t+1}^2(x_t)+\rho \hat{Z}_{t+1}(x_{t-1}(M), x_t),\,\, (1-\rho^2) \hat{\sigma}_{t+1}^2( x_t)\Big)
\end{split}
\end{equation*}
where
 $$
 \hat{\sigma}_{t+1}^2(x_t)=\frac{1}{M}\sum_{m=1}^M e^{x_t(m)},\quad \hat{Z}_{t+1}(x_{t-1}(M), x_t)=\sum_{m=1}^M e^{\frac{x_t(m)}{2}}\big(W^X_{t+m\delta}-W^X_{t+(m-1)\delta}\big).
 $$
Note that $W^X_{t+m\delta}-W^X_{t+(m-1)\delta}$ depends on $(x_{t-1}(M), x_t)$
through \eqref{eq:diff_X}. To complete the model we take for $\M_0(\dd x_0)$,
the initial distribution of process $\{X_t\}$, the density of the
$N\big(\mu^X,\omega^2/(2\kappa)\big)$ distribution.
 
We set the parameters of the model  to their estimated values for the daily return data on the closing price of the S\&P 500 index from 5/5/1995 to 4/14/2003 \cite{Chib2004} and simulate observations $\{Y_t\}_{t=0}^T$ using the discretized model \eqref{eq:diff_X}-\eqref{eq:diff_Y} with $M=20\,000$. The number of observations $T$ is set to $4\,000$.

Below we compare SMC with  SQMC based on the forward construction  and on the Brownian bridge construction of Brownian  paths. In both cases, SQMC is implemented using as input a nested scrambled  \cite{Owen1995} Sobol' sequence.  The performance of these three algorithms is compared, for $t=1,\dots,T$, for the estimation of (1) the filtering expectation $\mathbb{E}[X_t| Y_{0:t}]$ and (2)  of the log-likelihood function $\log(L_t)$.

\begin{figure}[!h]
\centering
\includegraphics[scale=0.32]{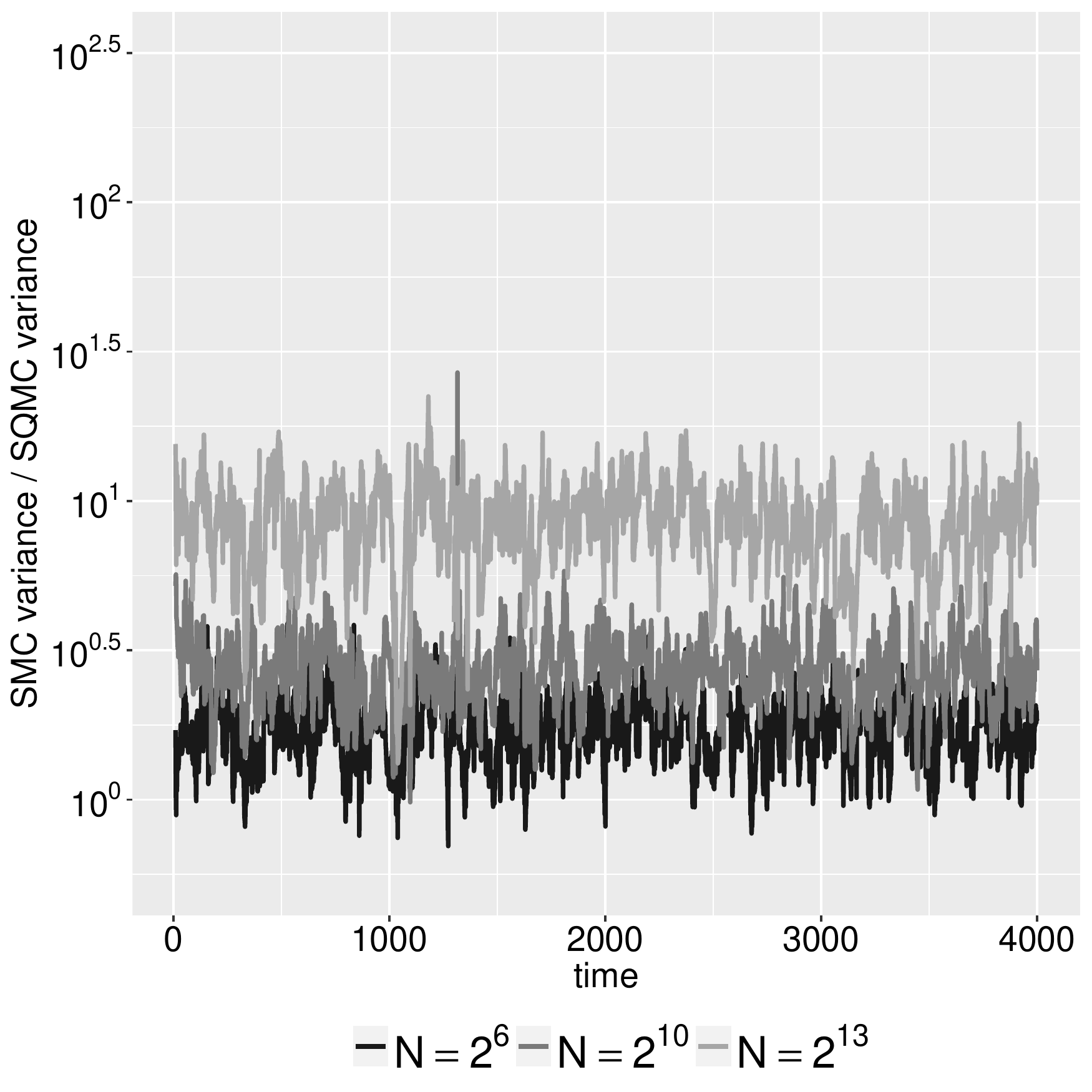}\includegraphics[scale=0.32]{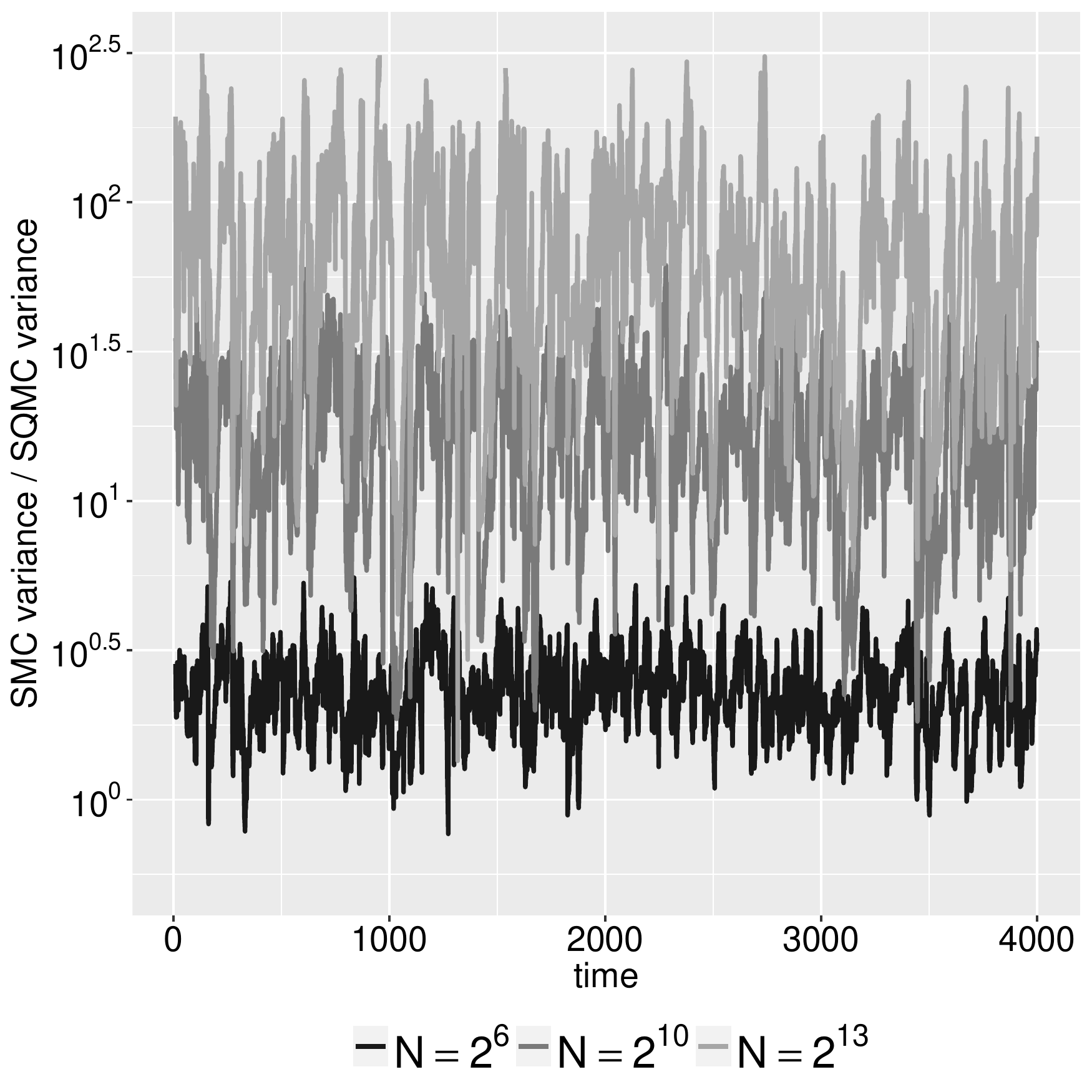}

\includegraphics[scale=0.32]{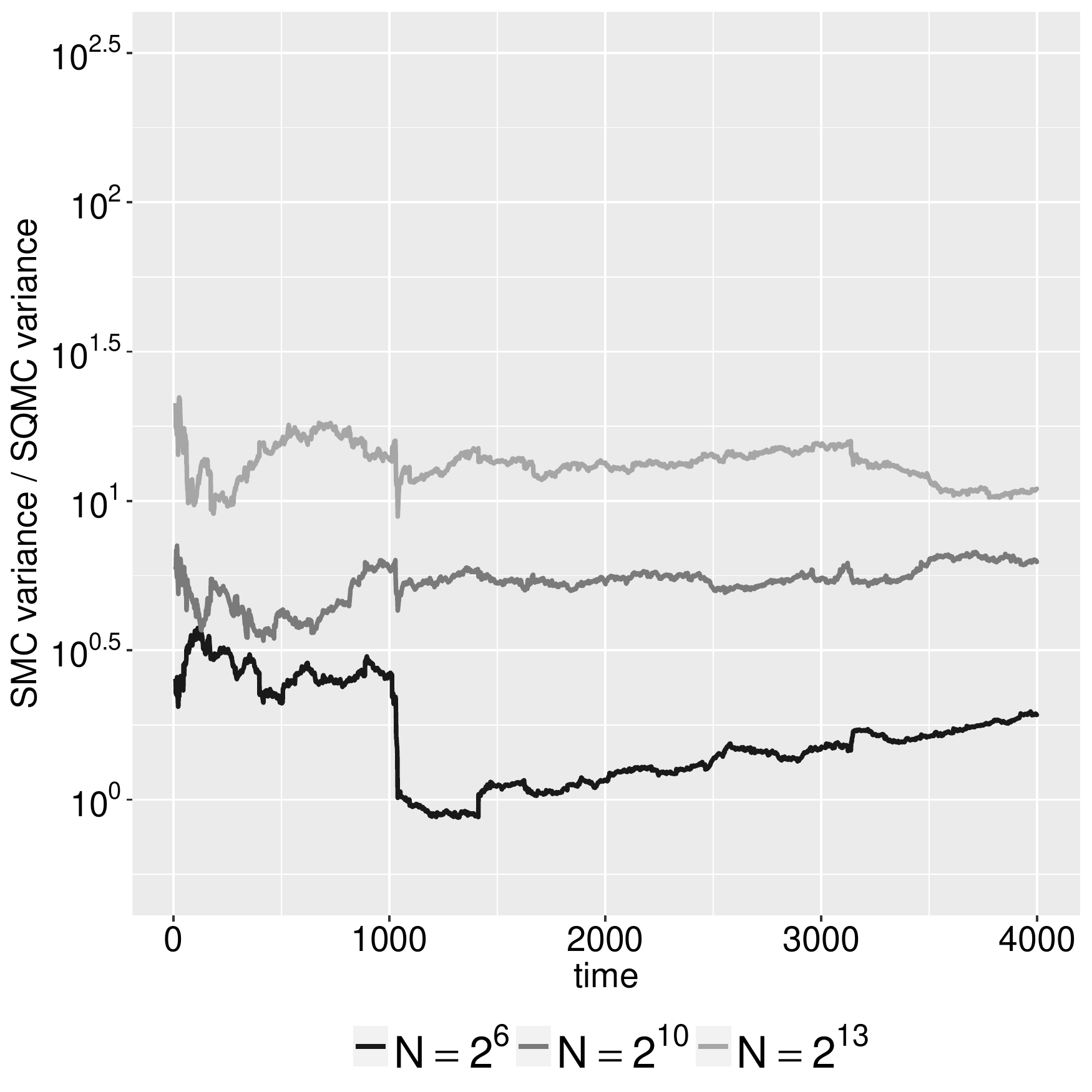}\includegraphics[scale=0.32]{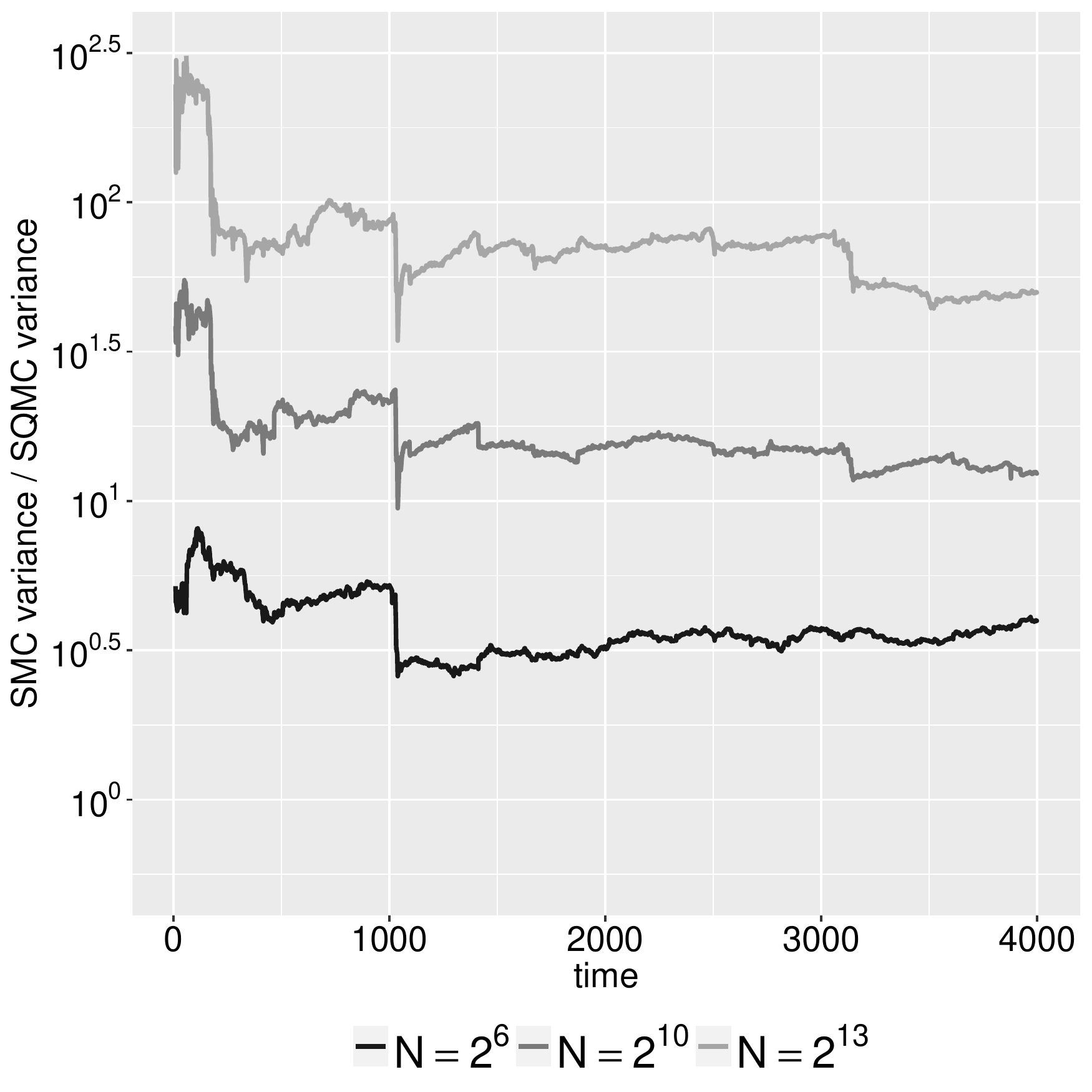}
\caption{Estimation of $\mathbb{E}[X_t|Y_{0:t}=y_{0:t}]$ (top plots) and of
$\log p(y_{0:t})$ for $t\in\{0,\dots,T\}$ and for different values of $N$. SQMC
is implemented  with the forward construction (left plots) and with the
Brownian Bridge construction of Brownian paths (right plots), and
$M=5$\label{fig:BBvsNormal}}
\end{figure}

Figure \ref{fig:BBvsNormal} shows the ratio of the SMC variance over the SQMC variance for the two alternative implementations of SQMC. Results are presented for a discretization grid of size $M=5$ and for different number of particles $N$. Two observations are worth noting from this figure. First, the two versions of SQMC outperform SMC in terms of variance. Second, the variance reduction is much larger with the Brownian bridge construction than with the forward construction of Brownian paths, as expected from the discussion of the previous subsection. Note that for both versions of SQMC the ratio of variances increases with the number of particles, showing that SQMC converges faster than the $N^{-1/2}$ Monte Carlo error rate.

\begin{figure}[!h]
\centering
\includegraphics[scale=0.33]{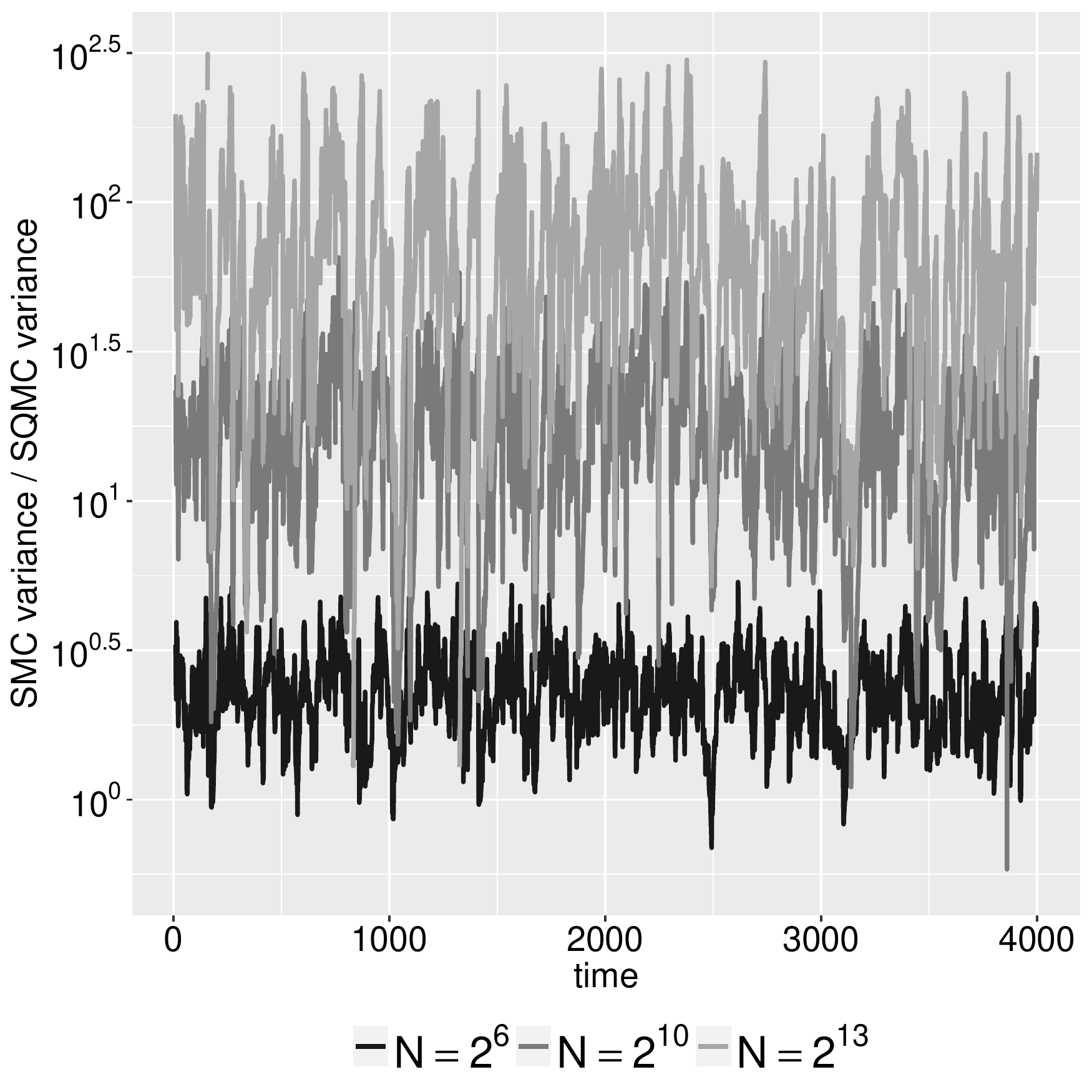}\includegraphics[scale=0.33]{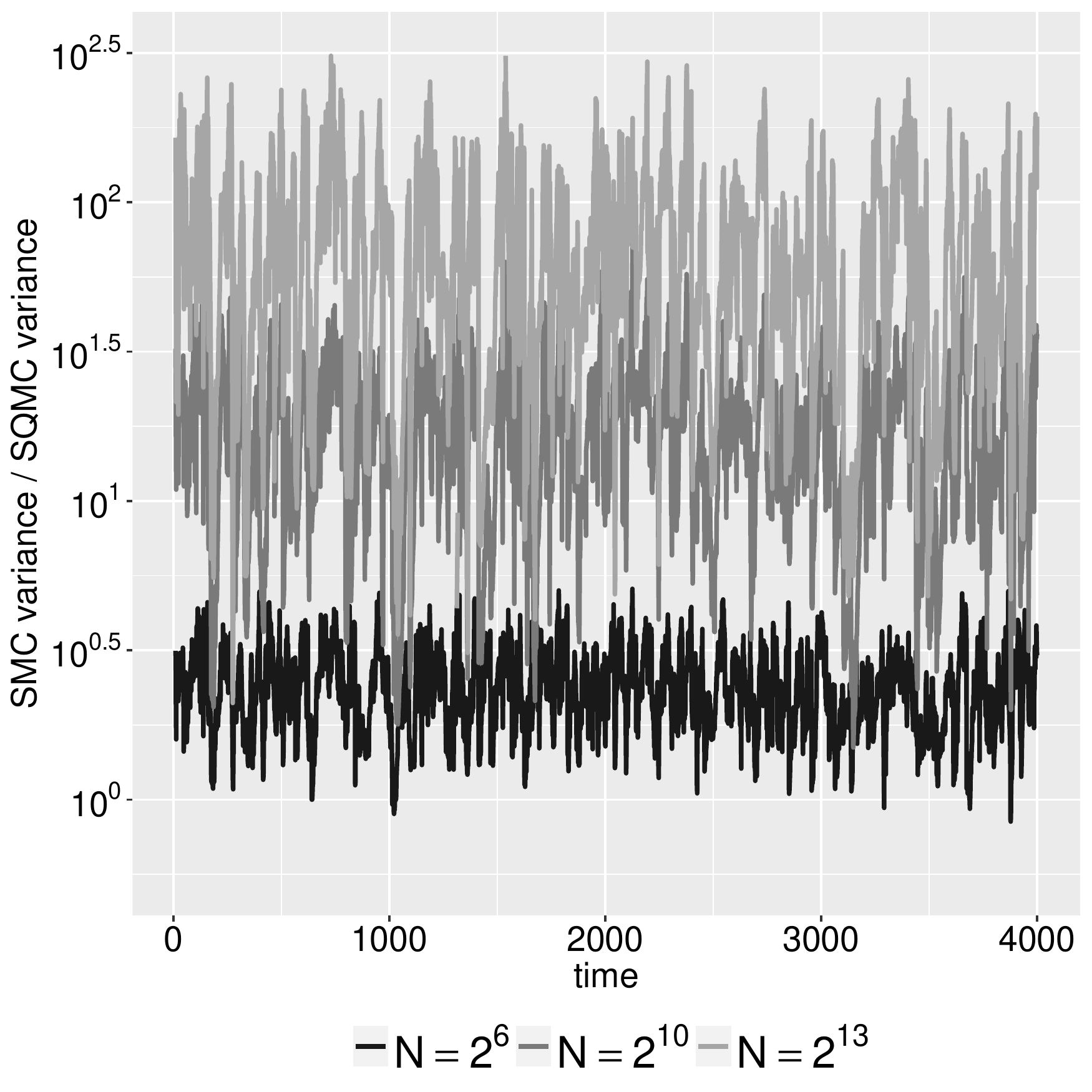}
\includegraphics[scale=0.33]{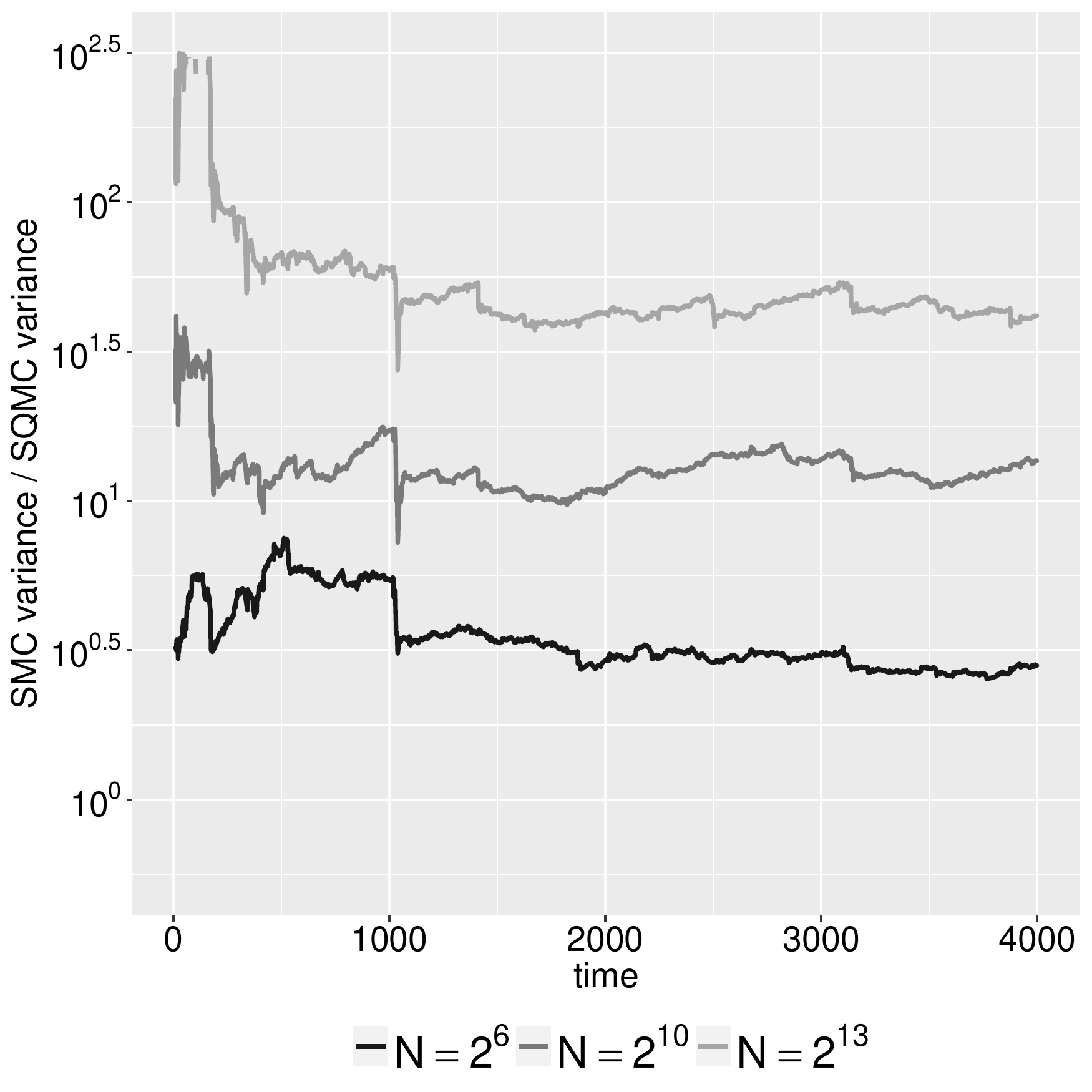}\includegraphics[scale=0.33]{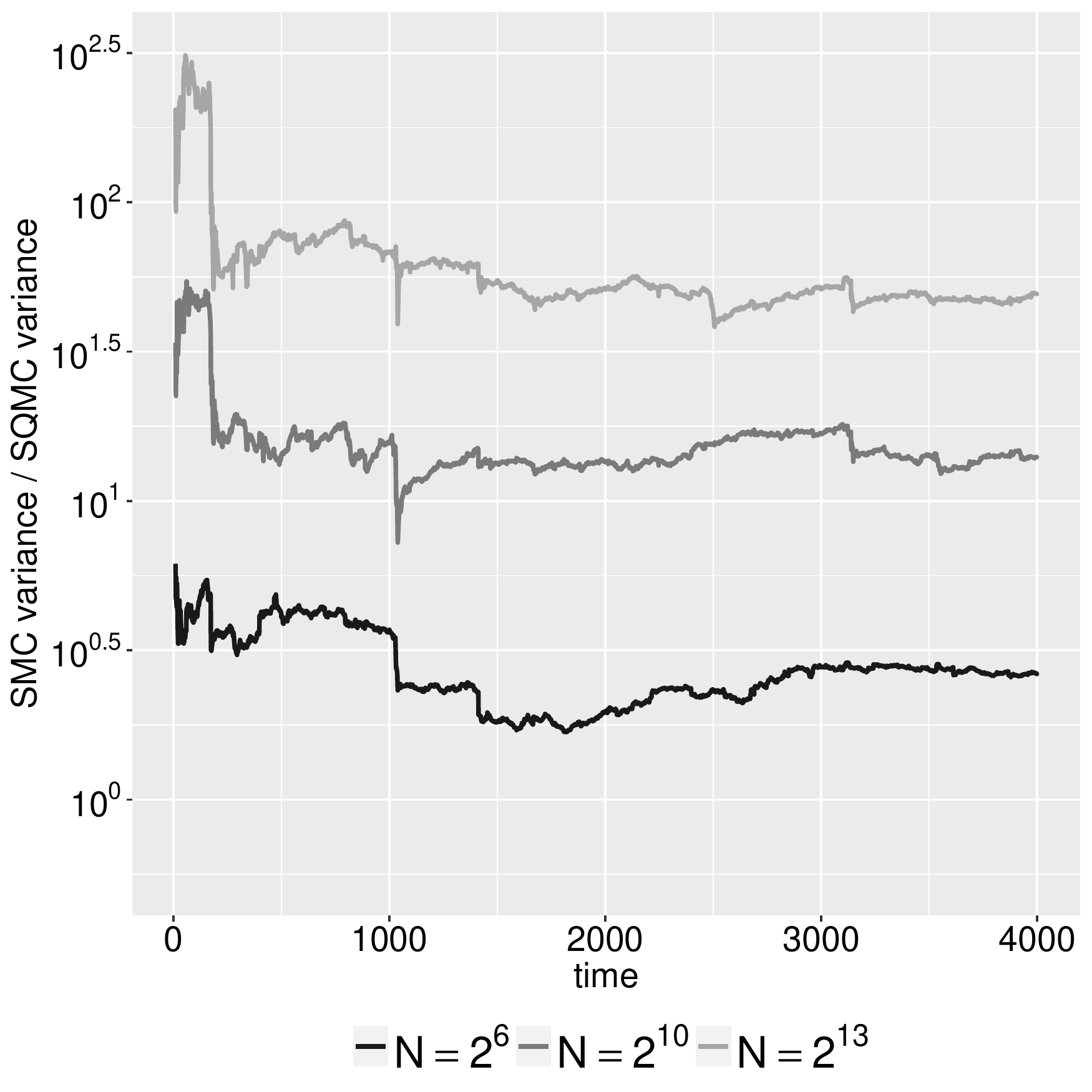}

\caption{Estimation of $\mathbb{E}[X_t|Y_{0:t}=y_{0:t}]$ (top) and of $\log
p(y_{0:t})$ for $t\in\{0,\dots,T\}$ and for different values of $N$. SQMC is
implemented  with the  Brownian Bridge construction of Brownian paths. Results
are presented for $M=10$ (left plots) and for $M=20$.\label{fig:variousM}}
\end{figure}

In Figure \ref{fig:variousM} we perform the same analysis than in Figure \ref{fig:BBvsNormal} but now with $M=10$ and $M=20$ discretization steps. ($M=10$ is considered as sufficient for parameter estimation by \cite{Chib2004}.) Results are presented only for the Brownian bridge construction. Despite the large dimension of the QMC point set used as input, we observe that SQMC converges much faster than the $N^{-1/2}$ Monte Carlo error rate. In particular, we observe that the gains in term of variance brought by SQMC are roughly similar whatever the choice of $M$ is. As explained above, this observation suggests that  the effective dimension of the model remains low (or even constant in the present setting) even when the ``true'' dimension $M$ increases.

\newpage

\newpage

\newpage
\bibliographystyle{spmpsci}

\appendix

\section*{Resampling}

Algorithm \ref{alg:inverse_method} below takes as input $N$ sorted
points $u^{1}\leq\ldots\leq u^{n}$, and $N$ weights $W^{n}$, and
return as an output the $N$ values $\left(F^{N}\right)^{-1}(u^{n})$,
where $\left(F^{N}\right)^{-1}$ is the inverse CDF relative to CDF
$F^{N}(z)=\sum_{n=1}^{N}W^{n}\ind\{n\leq z\}$, $z\in\R$. Its complexity
is $O(N)$. 

To compute the inverse CDF corresponding to the empirical CDF of $N$
ancestors (as discussed in Section \ref{subsec:SQMC-dim-one}), i.e.
\[
F^{N}(x)=\sum_{n=1}^{N}W^{n}\ind\left\{ X^{n}\leq x\right\} 
\]
simply order the $N$ ancestors, and apply the same algorithm to the
sorted ancestors.

\begin{algorithm}
\caption{\label{alg:inverse_method}Resampling Algorithm (inverse transform
method)}
\textbf{Input:} $u^{1:N}$ (such that $0\leq u^{1}\leq\ldots\leq u^{N}\leq 1$, $W^{1:N}$ (normalised weights)
\\
\textbf{Output:} $a^{1:N}$ (labels in $1:N$)

\hspace{0.3cm} $s\gets0$, $m\gets0$


\hspace{0.3cm}\textbf{for} $n=1\to N$  \textbf{do}

\hspace{0.6cm} \textbf{repeat}

 \hspace{0.9cm}$m\gets m+1$
 
 \hspace{0.9cm}$s\gets s+W^{m}$

\hspace{0.6cm} \textbf{until} $s>u^{n}$

\hspace{0.6cm} $a^{n}\gets m$

\hspace{0.3cm} \textbf{end for}

\end{algorithm}

\end{document}